\documentclass{article}
\usepackage{amssymb,amsmath}
\usepackage{blkarray}
\usepackage{color}
\usepackage{soul}
\usepackage{cancel}
\usepackage{amsthm}
\usepackage{multirow}
\usepackage{booktabs,caption}
\usepackage[flushleft]{threeparttable}
\usepackage{imakeidx}

\newtheorem{theorem}{Theorem}
\newtheorem{corollary}{Corollary}
\newtheorem{lemma}{Lemma}
\newtheorem{definition}{Definition}

\newtheorem{example}{Example}
\newtheorem{remark}{Remark}
\newtheorem{convention}{Convention}

\usepackage[utf8]{inputenc}
\usepackage[T1]{fontenc}

\title{Semi-Density Matrices and Quantum Statistical Inference(Corrected and Augmented Version)}

\author{Ahmad Shafiei Deh Abad, Mohammad Shahbazi\\\\
School of Mathematics, Statistics and Computer Science,\\ University of Tehran, Tehran, Iran \\ a\_shafiei@ut.ac.ir, mshahbazi@ut.ac.ir}
\begin{document}
\maketitle

\section{Abstract}

In this paper inspired by the "Minimum Description Length Principle" in classical statistics, we introduce a new method for predicting the outcomes of performing quantum measurements and for estimating the state of quantum systems.

\section{Introduction}

Needless to say{\color{black}{,}} nowadays nearly all our physical knowledge is based on quantum theory. So an increasingly important problem is to characterize quantum systems and to obtain information about them. In the way of solving the problem, Quantum Statistical {\color{black}{I}}nference (QSI) is a unique tool. As we know, quantum statistical inference is the quantum version of classical statistical inference. To be more precise, quantum statistical inference enables us to obtain information about quantum systems by using outcomes of performing quantum measurements. The research subject was initiated in the middle of {\color{black}{the}} 1960s. The pioneers and the first researchers in the field are Holevo, Yuen, Kennedy, Belavkin{\color{black}{, etc.}}
Their research is summarized in [8] and [9].
However, these researchers did not consider the asymptotic aspects while the asymptotic theory is essential for the large sample case in statistics and concerning it, an elegant general theory has been established in classical statistical inference theory. In the middle of the 1980s, a different research direction has been started by Nagaoka (who is an expert in mathematical statistics and information geometry), which focused on the asymptotic theory. In the 1990s, several Japanese researchers (Fujiwara, Matsumoto, Ogawa, Hayashi) have been influenced by Nagaoka, and joined Quantum Statistical Inference. Hence, in the 1990s, by combining the mathematical formulation of quantum mechanics and mathematical statistics, these Japanese researchers obtained several good results in Quantum Statistical Inference. Especially, the Japanese researchers have deeply discussed its asymptotic aspects, which had not been studied in the earlier stage. Recently, Quantum Statistical Inference has drawn the attention of several European statisticians (Gill, Bandorff- Nielsen, Jupp, Ballester, etc.) who joined this research field. On the other hand, several different directions of this research area were started in Europe after the 1990s by physicists, Massar, Popescu, D’Ariano, Buzek, Keyl, Werner, Bagan, Baig, Gisin, Vidal, Latorre, Pascual, Tarrach etc. They were motivated by the foundations of physics [7].

Since then till now many researchers in different countries{\color{black}{ have}} conducted research into the subject and{\color{black}{ have}} extended it in different directions. Among other things{\color{black}{,}} QSI contains the subject matters,{\color{black}{ quantum estimation and quantum prediction}}, which will be considered in this paper. To treat these{\color{black}{ problems}} the only tool at our disposal is performing measurements. Since quantum theory is statistical in nature, we have to perform the same quantum measurement{\color{black}{ in}}  the same state of the quantum system many times. But{\color{black}{,}} as it is well-known{\color{black}{,}} after performing {\color{black}{a}} measurement on a quantum system the state of the system changes drastically. To overcome the difficulty, we usually assume that there{\color{black}{ are}} $n$ quantum systems described by the same Hilbert space $\mathbb H$ and prepared independently and identically in the same state $\rho$ (a density matrix on $\mathbb H$) and we perform the same quantum measurement on each of them. In this way{\color{black}{,}} we obtain a data set ${{{\color{black}{D}}}}={{\color{black}{(}}} x_{1} , x_{2} , ... x_{n}{{\color{black}{)}}}$. By quantum estimation we mean{\color{black}{ techniques}} enabling us to find an approximation of the state $\rho$ with the help of the data set ${{\color{black}{D}}}$ and by prediction we mean characterizing the probability of the outcome $ x_{n+1}$ given the previous outcomes $x\in{{{\color{black}{D}}}}$. An appropriate method to solve the problems is to choose a set $\cal M$ of density matrices on $\mathbb{H}$ containing $\rho$, called a quantum model and try to find the state $\rho$ by methods{\color{black}{,}} such as Maximum Likelihood Estimation (MLE). To be able to act in this way{\color{black}{,}} we have to parameterize the set $\cal M$ in a differentiable manner. Unfortunately, ML Estimation which has been used by several authors gives rise to overfitting\footnote{{\color{black}{the selection of an overly complex model that, while fitting observed data very well, predicts future data very badly.}}}. Moreover, in general{\color{black}{,}} we do not know whether the state $\rho$ is in the model $\cal M$ or not.  Inspired by the works of J. Rissanen [14], [15], [16], P. Grünewald [3], [4] and others on{\color{black}{ the}} Minimum Description Length Principle (MDL) in classical statistics, one of our goals in this paper is to remedy this difficulty. Their works on the use of 2-part codes [4] in MDL guided us to use sets of semi-density matrices in addition to quantum models and call them generalized quantum models (for more detail see the beginning of Section 4). As in classical MDL we base our work on universal sources associated with quantum models. We will show that in all interesting cases universal quantum sources exist. It will be evident that {\color{black}{the}} use of universal sources automatically protects against overfitting. Moreover, we prove different versions of{\color{black}{ the}} consistency theorem showing that when the state $\rho$ is in the chosen model $\cal M$, the selected universal quantum source is asymptotically equivalent to it.

The organization of the paper is as follows:

In Section 3 we introduce the notion of Q-projection\index{Q-projection} which in this work will {\color{black}{act}} as projective quantum measurement and all the rest of this work are based on it. In Section 4 after some explanations about the MDL principle and the way we have gone through to quantize the most important notions involved in MDL, we will define fundamental concepts{\color{black}{,}} such as (generalized) quantum model{\color{black}{s}}, universal quantum source{\color{black}{s,}} which is the core concept of this work, quantum source and quantum strategy\index{Quantum strategy}{\color{black}{.}} We will {\color{black}{also}} prove some important facts about them. At the end of the same section{\color{black}{,}} we introduce the notion of good quantum estimator and a large class of them. Section 5 is about quantum prediction\index{Quantum prediction} and quantum estimation\index{Quantum estimation}. In Section 6 we will introduce the notion of consistency and prove some theorems about it. {{\color{black}{In Section 7, we give two examples that indicates the efficiency of this method.}}}

We emphasize that with the help of trace function, one can reduce the problems treated here to problems in the classical MDL methods and solve them classically. But in doing this the operator nature of important concepts like universal quantum source associated with quantum models\index{Quantum models}, quantum strategy and conditional density matrix\index{Conditional density matrix} conditioned on  density matrix will be lost. Even worse, one cannot understand that these concepts are operators. Moreover, treating the problems in the realm of operator theory are more natural and simpler. In the same vein, nearly all notations, definitions and conventions used in the paper is directly inspired by their classical counterparts in [4]. So that comparison of classical and quantum frameworks should be straightforward.

It is necessary to mention that the proof of Theorem 2 of [17] is incomplete and there are other errors in it. All of them are corrected in this paper.

\section{Q-Projection}

Given a separable Hilbert space $\mathbb H$, in general infinite dimensional, with inner product $\langle \cdot | \cdot\rangle$, the set $\{ |k\rangle | k\in {\mathbb N} \}$ will denote an orthonormal basis of $\mathbb H$ and its dual basis will be denoted by the set $\{ \langle k| \vert k\in {\mathbb N} \}$. The set of all bounded operators (resp. self-adjoint bounded operators) on $\mathbb H$ will be denoted by $B(\mathbb H)$ (resp. by $B_{H}(\mathbb H)$) and the set of all positive operators (resp. density matrices) on $\mathbb H$ will be denoted by $B_{+}(\mathbb H)$ (resp. by $D(\mathbb H)$). Finally, the Hilbert space generated by trace class operators of $\mathbb H$ with the following inner product will be denoted by $B_{T}{(\mathbb {H})}$,

$$<S | T>_{T}=Tr(T^{\star}{S}) , for all T,S\in {B_{T}}{(\mathbb{H})}$$

with associated norm
${\Vert T\Vert}_{T}=\sqrt{{\rm Tr}(T^{\ast}T)}$.

A positive operator $T$ is called a semi-density matrix if $Tr(T)\leq{1}$ and it is called a density matrix if $Tr(T)={1}$. The mapping which sends each nonzero semi-density matrix $T$ to its associated density matrix $\frac{T}{Tr(T)}$ will be denoted by $\omega$.The collection of all complete sets of mutually orthogonal (minimal) projections $ P= \{ p_{1}, p_{2}, ...\}$ on $\mathbb{H}$, {with $\sum_{n}p_n=1$ (completeness),} will be denoted by ${\pi}({\mathbb{H}})$ (${\pi }_{0}({\mathbb{H}})$).

{\color{black}{Let $P= \{ p_{1}, p_{2}, ...\}$ and $Q= \{ q_{1}, q_{2}, ...\}$ be elements of ${\pi }({\mathbb{H}})$.Then, the set $\{p_{i}q_{j} | i , j \in{\mathbb{N}} \}-\{0 \}$ will be denoted by $PQ$ and will be called the combination of $P$ and $Q$. We say that $P$ and $Q$ commute if $PQ=QP.$ In this case clearly $PQ\in{\pi}({\mathbb H})$. More generally, a subset $\mathbb {S}$ of ${\pi }({\mathbb{H}})$ is called commutative, if any two elements of it commute. Let
${\mathbb{P}}=\{P_{1} , P_{2} , ... ,P_{k} \}$ be a finite subset of $ {\pi}({\mathbb{H}})$. The combination of elements of ${\mathbb{P}}$ is
$${\Pi}_{i=1}^{k}{P_{i}} =\{ {\Pi}_{i=1}^{k}{x_{i}} |x_{i}\in P_{i} , i=1 , 2 , ... , k \}-\{0 \}.$$
When ${\mathbb P}$ is commutative, ${\Pi}_{i=1}^{k}{P_{i}}\in {\pi }({\mathbb H})$.
}}

\begin{definition}
Assume that $(X_{j})_{j\in{J}}$ is a family of subsets of a nonempty set $X$ and for each $j\in J$,
there exists {\color{black}{ $I_j\subset J$}}  such that $X-{X_{j}}=$ {\color{black}{ ${\cup}_{i\in I_j}{X_{i}}$}}. {{\color{black}{Then,}}}

\begin{enumerate}
\item {\color{black}{For each $I \subset J$,}} we say that the set $Y={\cup}_{i\in {I}}{X_{i}}$ is a maximally connected\index{Maximally connected} union of the family $(X_{j})_{j\in{J}}$ if it{\color{black}{ satisfies}} the following {\color{black}{ conditions}}:
\begin{enumerate}
\item  For each proper subset $K$ of $I$,
$${{\cup}_{k\in K}{X_{k}}}\cap({ {\cup}_{i\in I-K}{X_{i}}})\neq{\{ \}}.$$

\item {\color{black}{$$Y{\cap}({\cup}_{i\in{J-I}}{X_{i}})=\{\}.$$
The set of all maximally connected unions of the family $(X_{j})_{j\in{J}}$ will be denoted by ${\wedge}_{j\in J}{X_{j}}$. Clearly, ${\wedge}_{j\in J}{X_{j}}$ is a partition of $X$.}}
\end{enumerate}

\item {\color{black}{For each $I \subset J$, the}} non-empty subset $Z={{\cap}_{i\in I}X_{i}}$ of $X$ will be called a minimally connected intersection of the family
$(X_{j})_{j\in J}$, if

{\color{black}{$$Z{\cap}({{\cup}_{j\in{J-I}}{X_{j}}})=\{\}.$$}}

The set of all minimally connected intersections of the family $(X_{j})_{j\in{J}}$ is evidently a partition of $X$ and will be denoted by ${\vee}_{j\in J}{X_{j}}$.
\end{enumerate}
\end{definition}
Now assume that $X$ is an arbitrary non-empty set. Let the set of all partitions of X  be denoted by $\mathfrak{P}(X)$. Let $\underline{P}$ and $\underline{Q}$ be in $\mathfrak{P}(X)$. We say that $\underline{Q}$ is finer than $\underline{P}$ and we write $\underline{P}\preceq \underline{Q}$, if each elements of $\underline{P}$ is the union of some elements of $\underline{Q}$. It is evident that the set $\mathfrak{P}(X)$ with the order relation $\underline{P}\preceq \underline{Q}$ is a partially ordered set. Assume that ${\mathbb{P}}=\{\underline P_{k}|k\in K \}\subseteq{\mathfrak{P}(X)}$ is a set of partitions of the set $X$. Let $\cup_{k\in K} \underline{P}_k=\{X_j\vert j \in J\}$. Clearly $X={\cup}_{j\in J}X_{j}$ {\color{black}{and the family $(X_{j})_{j\in{J}}$ of subsets of the set $X$ satisfies the conditions of Definition 1.}} It is easy to see that for each partition $\underline {P_{k}}\in {\mathbb{P}}$ we have

$${\wedge}_{j\in {J}}{X_{j}}\preceq\underline{P_{k}}\preceq{\vee}_{j\in{J}}{X_{j}}.$$

Let partitions $\underline P$ and $\underline Q$ of the set $X$ be such that for all $k \in K$ we have $\underline Q\preceq{\underline P_{k}}\preceq \underline P.$ Then, it is straightforward to see that for each $k\in K$

$$\underline{Q}\preceq{\wedge}_{j\in J}{X_{j}}\preceq\underline{P}_{k}\preceq{\vee}_{j\in J}{X_{j}}\preceq \underline{P}.$$

Therefore, ${\wedge}_{j\in J}{X_{j}}$ (resp. ${\vee}_{j\in J}{X_{j}}$ ) is the greatest lower bound (resp. the least upper bound ) of the partially ordered set $\mathbb P$ and will be denoted by

${\wedge}_{k\in K}{\underline P_{k}}$ (resp. ${\vee}_{k\in K}{\underline P_{K}}$).

\begin{example} Let $X=\{ a,b,c,d,e,f,g,h,i,j,k \}$ and let
$X_{1}=\{a,b\}$ , $X_{2}=\{b,c\}$ , $X_{3}=\{c,d\}$ , $X_{4}=\{d,a\}$ , $X_{5}=\{e,f,g\}$ ,
$X_{6}=\{f,g,h\}$ , $X_{7}=\{g,h,i\}$ , $X_{8}=\{h,i,j\}$ , $X_{9}=\{i,j,k\}$ , $X_{10}=\{j,k,e\}$ ,
$X_{11}=\{k,e,f\}$.

Clearly, the set $X$ and its subsets $X_{j} , j=1,2, ... , 11$ satisfy the conditions of above definition and we have $$\vee_{1\leq{j}\leq{11}}{X_{j}}=\{\{a\} , \{b\} , \{c\} , \{d\} , \{e\} , \{f\} , \{g\} , \{h\} ,
\{i\} , \{j\} , \{k\}\},$$ $$\wedge_{1\leq{j}\leq{11}}{X_{j}}=\{\{a,b,c,d\} , \{e,f,g,h,i,j,k\}\}.$$

\end{example}

\begin{definition}
\label{def4.1} Let $ P$ and $Q$ be in ${\pi }({\mathbb H})$. We say that $P$ is \textbf{finer} than $Q$, and we write $Q\preceq P$ if
{\color{black}{$PQ=P.$ In this case $QP=(PQ)^{\star}=P.$}}
\\
We say that $Q$ and $P$ are \textbf{consistent} if they have a common upper bound with respect to this order relation. More generally, a subset $ A\in{\pi(\mathbb{H})}$ is called consistent if it has an upper bound. Then clearly any subset of $A$ is also consistent. we say that a consistent set $A$ is maximally consistent if there is no consistent subset B of $\pi({\mathbb H})$ such that $A\subsetneqq B$.

\end{definition}

\begin{lemma}
\label{lem4.2.1} Let $\mathbb{P}=\{P_{k}\in{{\pi}(\mathbb H)} \vert k\in K\}$. Then
\begin{enumerate}

\item If the set is consistent it has a least upper bound\index{least upper bound} and a greatest lower bound\index{greatest lower bound}.

\item If the set is {\color{black}{finite and commutative, then it is}} consistent.
\end{enumerate}
\end{lemma}

\noindent{\bf{Proof}.}\quad
\begin{enumerate}

\item Assume that the set $\mathbb{P}$ is consistent then it has an upper bound $R=\{ r_1, r_2, \cdots\}$ which is a complete set of mutually orthogonal projections of $\mathbb{H}$.
Let $Q\in\mathbb{P}$. By definition $Q \preceq{R}$. Let $q\in{Q}$ and let $R_q$ be the sum of all elements $r\in{R}$ such that $q\geq{r}$. i.e. $qr=r.$ Clearly $R_q^2=R_{q}\neq 0$ and $qR_q=R_qq=R_q$, since $rq=r$ for all $r\in R_q$. Therefore
$q-R_q$ is a projection and if
$q-R_{q}=q(I_{\mathbb{H}}-R_{q})\neq{0}$,
then there exists $r\in R$ such that $q\geq r$ and $rR_q=0$ 
which is a contradiction. Hence, $q=R_{q}$.
Therefore for each $Q\in\mathbb{P}$, each $q\in Q$ is the sum of some elements of $R$.

Let the order preserving mapping $Q\rightarrow \underline Q$ from $\pi(\mathbb{H})$ into $\mathfrak{P}(R)$ be defined as follows,
for each $q\in Q$, $q\rightarrow \underline q$, where $\underline q$ is the set of all summands of the projection $q$.
Notice that $q$ is the sum of some elements of $R$.
Now it is clear that under this mapping we have the following bijective maps.

$$\wedge_{k\in{K}}{ P_{k}}\rightarrow \wedge_{k\in{K}}{\underline{P_{k}}}$$
$${\vee}_{k\in{K}}{ P_{k}}\rightarrow {\vee}_{k\in{K}}{\underline{P_{k}}}$$
we have seen above that ${\vee}_{k\in{K}}{\underline{P_{k}}}$ (resp. $ {\wedge}_{k\in{K}}{\underline P_{k}}$) is the least upper bound (resp. the greatest lower bound) of the set $\{{\underline{P}}_{k} , k\in{K}\}$. Therefore,
$\wedge_{k\in K}{ P_{k}}$(resp.$ {\vee}_{k\in K}{ P_{k}}$) is the greatest lower bound (rep. the least upper bound) of $\mathbb{P}$.

\item Assume that the set $\mathbb{P}$ is {\color{black}{finite and}} commutative. Then, ${\vee}_{k\in K}{ P_{k}}={\Pi}_{k\in K}P_{k} $. Therefore, $\mathbb{P}$ is consistent.
\end{enumerate}

$\hfill\blacksquare$

\begin{definition}
Let $T\in B(\mathbb H)$ and $Q = \{ q_{1}, q_{2}, ... \}\in {\pi(\mathbb H)}$. Then The element
$${T_{Q}}={\sum_{n}{q_{n}{T}q_{n}}},$$
will be called the \textit{$Q$-projection} of $T$ (see also [1]).
The set of all $Q$-projections of elements of $B(\mathbb H)$ will be denoted by $B_{Q}(\mathbb H)$ and for each $q\in Q$, $B_{q}{(\mathbb{H})}=\{qTq\vert T\in{B(\mathbb{H})}\}.$
\end{definition}
The set $B_{Q}(\mathbb H)$ is a complex subspace of the ${C}^{\ast}-algebra$ $ B(\mathbb H)$, and the mapping $\bar {Q}$ from $B(\mathbb H)$ into $B_{Q}(\mathbb H)$ defined by
$\bar {Q}(T) := T_{Q}$ is a projection. For $T$ and $S$ in $B(\mathbb{H})$ and $Q\in \pi (\mathbb{H})$ we have $ (T_{Q}S_{Q} )_{Q} =T_{Q}S_{Q}.$Therefore $B_{Q}(\mathbb{H})$ is a unital ${C}^{\ast}-$subalgebra of $B(\mathbb{H})$. If $Q\in \pi_0(\mathbb{H})$ then evidently $B_Q(\mathbb{H})$ is commutative.

\begin{lemma}
\label{lem4.2}\hfill

\begin{enumerate}
\item The mapping $\bar {Q}$ is trace preserving.
\item If $T$ is self-adjoint, then $T_{Q}$ is also self-adjoint.
\item A necessary and sufficient condition for $T$ to be positive is that for each $Q\in{\pi}(\mathbb{H})$, $T_{Q}$ be positive.
\item Let $Q \in \pi_0({\mathbb{H}})$ and $T \in B(\mathbb{H})$ be arbitrary. Then, $T_Q$ is always normal.
\end{enumerate}
\end{lemma}

\noindent{\bf{Proof}.}\quad
\begin{enumerate}
\item $
{\rm Tr}(T_{Q})
=\sum_{n=1}^{\infty}{\rm Tr}(q_{n}Tq_{n})=\sum_{n=1}^{\infty}{\rm Tr}(q_{n}T)
=Tr(T),
$
since the sets of projections $Q\in\pi(H)$ are complete.
\item

{If $T=T^\ast$, then  evidently $(T_Q)^\ast=T_Q$.}

\item {Let $T\geq 0$; then, for each $q\in Q$ , $qTq\geq 0$. So that
for each $Q\in\pi(\mathbb{H})$ , $T_{Q}\geq 0$. Vice versa, if $T_{Q}\geq 0$ for each $Q\in {\pi}(\mathbb{H})$, then, for each vector $\vert v\rangle\in {\mathbb{H}}$, $Tr(\vert v \rangle\langle v\vert T)=\langle v \vert T\vert v\rangle\geq 0$, since any such $\vert v \rangle\langle v\vert$ belongs to some $Q\in\pi(\mathbb{H})$, $T\geq 0$.}
\item Since in this case $B_{Q}({\mathbb{H}})$ is a commutative algebra, the proof is clear.

$\hfill\blacksquare$
\end{enumerate}
\begin{corollary}
\label{cor4.1} The restriction of the mapping $\bar {Q}$ to $D(\mathbb H)$ is a convex map from $D(\mathbb H)$ onto $D_{Q}(\mathbb H)$.
\end{corollary}

\begin{lemma}\hfill
\label{lem4.3}
\begin{enumerate}
\item The mapping ${\bar Q}:B(\mathbb H)\longrightarrow{B_{Q}(\mathbb H)}$ is continuous.
\item The mapping ${\bar Q}:B_{T}{(\mathbb H)}\longrightarrow{B_{Q}(\mathbb H)}$ is continuous in the $|| , ||_{T}$ topology.
\end{enumerate}
\end{lemma}

\noindent\textbf{Proof.}\quad
\begin{enumerate}
\item

Let $T\in B(\mathbb{H})$ be a self-adjoint element of $B(\mathbb H)$.
Then, $\Vert T_{Q}\Vert $ is equal to its spectral radius $r$. Let $q\in Q$ and let $\Vert qTq\Vert =r$.
Then
$$
|| T_{Q}||=||\sum_{n} {q_{n}Tq_{n}}||=|| qTq||{ \leq||T||}
$$
Since any $T\in B{(\mathbb H)}$ can be written as a combination of two self adjoint elements $\bar{Q}$ is continuous.
\item Let $T\in B_{T}{(\mathbb H)}$. Then, for each $q\in Q$ , $qTqqT^{\ast}q\leq {qTT^{\ast}q}.$ Therefore,
$(T_{Q})^\ast T_{Q}=((T^\ast)_{Q})T_Q\leq(T^\ast T)_{Q}$.
Since ${\rm Tr}(T_{Q})={\rm Tr}(T)$,
$$||T_{Q}||_{T}=({Tr({T_{Q}}^{\star}{T_{Q}})})^{1/2}\leq{(Tr{((T}^{\star}{T})_{Q})})^{1/2}
=(Tr({T}^{\star}{T}))^{1/2}=||T||_{T}.$$
\hfill$\blacksquare$

\end{enumerate}
\bigskip

\begin{lemma}
\label{lem4.4} For each element $T\in{ B(\mathbb{H})}$ and each $Q= \{ q_{1}, q_{2}, ... \} \in {\pi( {\mathbb{H}}})$ we have:
\begin{enumerate}
\item $T=T_{Q}$ if and only if for each $q\in Q$ we have $qT=Tq$.
\item Let $S=S_Q$ and for all $q\in Q$ , $qSq=qTq$. Then, $S= T_{Q}.$
\item Let $T$ be a normal operator and $f$ be a continuous function defined on a neighborhood of the spectrum of $T$. If $T=T_{Q}$ then $f(T)=(f(T))_{Q}.$
\end{enumerate}
\end{lemma}

\noindent\textbf{Proof.}\quad
\begin{enumerate}
\item Assume that {$T=T_{Q}=\sum_nq_nTq_n$.
Then, for each $q_n\in Q$ we have
$$
{q_{n}}T={q_{n}}T_{Q}={q_{n}}T{q_{n}}=T_{Q}{q_{n}}=T{q_{n}} .
$$
Conversely, if for each $q_n\in Q$, $Tq_n=q_nT$, then, completeness of $Q$ yields
$$
T_{Q}=
\sum_{n}{q_{n}}T{q_{n}}=\sum_{n}{q_{n}}T=T.$$}
\item {By hypothesis,
$
S=S_{Q}=
\sum_nq_nSq_n=\sum_nq_nTq_n=T_Q
$.}
\item The proof is a consequence of {point $1$} and of functional calculus.
\end{enumerate}

$\hfill\blacksquare$
\bigskip

\begin{lemma}
\label{lem4.6_1} Let $T\in {B(\mathbb H)}$ and $P, Q\in {\pi ({\mathbb H})}$. If {$ P\succeq Q $} then:

1) $T_{P}=(T_{Q})_{P}=(T_{P})_{Q}.$

2) $Ker(\bar{Q})\subset Ker(\bar{P})$

\end{lemma}

\noindent\textbf{Proof.}\quad It is clear that for each element $p\in P$ there exists exactly one element $q_{0}\in Q$ such that
$q_{0}p=pq_{0}=p$ and for other elements $q\in Q$ we have $qp=pq=0$. So
$$p(T_Q)p=p(\sum_{q\in Q}qTq)p=pq_0 Tq_0 p=pTp$$
Therefore,
$$(T_Q)_P=\sum_{p\in P}pT_Q p =\sum_{p\in P} pTp=T_P$$
On the other hand for each $q \in Q$ and each $p \in P$ we have

\begin{equation}
\begin{array}{rl}
qT_{P}&=q{\sum}_{p\in P}pTp={\sum}_{p\in P}qpTp \\
&={\sum}_{p\in P\vert qp\neq{0}}pTp={\sum}_{p\in P}pTpq\\
&=T_{P}q.
\end{array}
\end{equation}

Therefore $T_{P}=(T_{Q})_{P}={(T_{P})}_{Q}.$

Since $\bar{P}(T)=T_P=(T_Q)_P=\bar{P}(\bar{Q}(T))$, the proof of the second part is clear.

$\hfill\blacksquare$
\bigskip

Let S and T be in $B(\mathbb{H})$. Then, in general $ST\neq TS$. But for all $Q\in \pi_0(\mathbb{H})$, $S_QT_Q=T_QS_Q$. This fact motivates the following definition.

\begin{definition}
\label{def4.2} Let $R$ be an $n$-ary relation on $B(\mathbb{H})$. We say that $R$
is \textit{weakly true}\index{weakly true} if, for each $Q \in \pi_0(\mathbb{H})$, ${\bar{Q}^{n}}(R)$ is true, where ${\bar{Q}^{n}}(R)$ is the image of $R$ under ${\bar{Q}^{n}}$, the natural extension of $\bar{Q}:B(\mathbb{H})\rightarrow B_{Q}(\mathbb{H})$ to $\bar{Q}^{n}:(B(\mathbb{H}))^{n}\rightarrow (B_{Q}(\mathbb{H}))^{n}$
\end{definition}
\begin{remark}
Any two elements of $B(\mathbb H)$ always weakly commute. For some relations, being true or weakly true are equivalent. For example, if $T\geq{S}$ then clearly, this relation is weakly true. \\
Conversely, Assume that for each $Q \in \pi_0(\mathbb{H})$, $T_{Q}\geq{S_{Q}}$ therefore for each minimal projection $q$, $qT{q}\geq{qS{q}}$. Since for each vector $v\in{\mathbb H}$ the projection $|v><v|$ is contained in some $ Q \in \pi_0(\mathbb{H})$ we have $\langle v\vert T-S\vert v\rangle\geq{0}.$ Therefore,
$T-S\geq{0}$.
\end{remark}
The relation weakly equal will be denoted by $=^{w}$.

\begin{lemma}
\label{lemNEW4.6_2} Let $\mathbb{H}$ be a separable Hilbert space and let $T\in B(\mathbb H)$ be a nonzero operator. Then:

1) If $T$ is invertible then, it is weakly invertible.

2) If $T$ is normal and weakly invertible then, it is invertible.
\end{lemma}

\noindent{\bf{Proof}.}\quad

1) Let $T$ be invertible, and let $Q= \{q_{1}, q_{2}, ... q_{n}, ...\}$ be an arbitrary element of ${\pi}_{0}(\mathbb H)$. We are going to prove that the operator $T_{Q}$ is invertible. Clearly, there exists an orthonormal basis ${\bf{b}}=\{|e_{1}>, |e_{2}>, ... |e_{n}>, ...\}$ for $\mathbb H$ such that for each $n\in{\mathbb N}$ we have $q_{n}=|e_{n}><e_{n}|$. Let $v={\sum}_{n=1}^{\infty}{{\lambda}_{n}{|e_{n}>}}$ be an arbitrary element of $\mathbb H$. Then
$$T_{Q}{v}={\sum}_{n=1}^{\infty}{q_{n}{T}q_{n}}{({\sum}_{n=1}^{\infty}{{\lambda}_{n}{|e_{n}>}})}=$$ $$
{\sum}_{n=1}^{\infty}{{\lambda}_{n}q_{n}T|e_{n}>}={\sum}_{n=1}^{\infty}{({\lambda}_{n}<e_{n}|T|e_{n}>)
|e_{n}>}.$$

Since, $T$ is invertible and $v$ is not $0$, $T_{Q}{v}$ is not $0$. Therefore, $T_{Q}$ is invertible and $T$ is weakly invertible.

2) Assume that $T$ is normal and weakly invertible. We are going to prove that $T$ is invertible. Suppose that $|e_{n}>$'s are eigenvectors of $T$ and $q_{i}$'s are their spectral projections. Since
$T$ is weakly invertible. $T_{Q}$ is invertible. But $T=T_{Q}$. Therefore. $T$ is invertible.

\begin{lemma}
\label{lem4.7} Let $T=T_{Q}$ be an invertible element of $B(\mathbb H).$ Then $T^{-1}=(T^{-1})_Q.$
\end{lemma}

\noindent{\bf{Proof}.}\quad From Lemma~\ref{lem4.4} and the fact that $qT=Tq$ implies $q=TqT^{-1}$, it follows that $T^{-1}q=qT^{-1}$.$\hfill\blacksquare$
\medskip

Let $T=T_{Q}$ be a normal operator. Then $T_{Q}$ is called a \textit{pseudo-spectral decomposition} of $T$. Clearly, for each $q\in {Q}$, $q(\mathbb H)$ is invariant under $T$.

\begin{lemma}
\label{lem4.9} Assume that $T_{P}$ is a pseudo-spectral decomposition of the operator $T$. Then for each $S\in {B(\mathbb H)}$, we have
$$(ST)_{P}=S_{P}T_{P}\quad\text{and}\quad (TS)_{P}= T_{P}S_{P}\\ , \\ Tr(TS)= Tr(T_{P}S_{P}).$$
\end{lemma}

\noindent{\bf{Proof}.}\quad We have $(ST)_{P}=(ST_{P})_{P}.$ Therefore, for each $p\in P$ we have
$p(ST)p=p(ST_{P})p=pSpTp=(pSp)(pTp).$ Therefore, $(ST)_{P}=S_{P}T_{P}$.
The proof of the second equality is the same. The third equality is evident.$\hfill\blacksquare$
\bigskip

{The previous lemmas lead to the following result.}

\begin{theorem}
\label{theo4.1} Let $Q$ be in $\pi(\mathbb H)$. Then
\begin{enumerate}
\item $B_{Q}(\mathbb H)$ is a unital ${C}^{\ast}$-algebra.
\item $B(\mathbb H)$ is a left and a right $B_{Q}{(\mathbb H)}$-module.
\item The mapping $\bar Q$ from $B(\mathbb H)$ into $B_{Q}(\mathbb H)$ is a $B_{Q}(\mathbb H)$-linear form.
\item A necessary and sufficient condition for $B_{Q}(\mathbb H)$ to be commutative is that $Q$ be a complete set of mutually orthogonal minimal projections.

\end{enumerate}
\end{theorem}

Let $\rho\in D(\mathbb{H})$ be a diagonal matrix. Clearly, we can consider $\rho$ as a classical probability distribution function. But if the density matrix $\rho$ is not diagonal we cannot interpret it in this way. The following definition serves to discriminate these two cases.

\begin{definition}
\label{def4.3} Let $\mathbb H$ be a separable Hilbert space and $Q\in{{\pi}_{0}(\mathbb H)}.$ The mapping ${\nu}:B(\mathbb H)\longrightarrow{\mathbb R}$ given by ${\nu}(T)=\Vert T-T_{Q}\Vert $ will be called
$Q$-quantum complexity of $T$. When ${\nu}(T)=0$, $T$ is called $Q$-classical and when $T_{Q}=0$, $T$ will be called $Q$-maximally non-classical.
The Von Nuemann entropy of $T_Q$ will be called the $Q$-Shannon entropy of $T$.

\end{definition}

\begin{example}
\label{ex4.1} Let $\mathbb H$ be a 2-dimensional Hilbert space with the standard basis $\{\vert 0\rangle ,\vert 1\rangle \}$. Let $X$, $Y$, $Z$ be Paoli density matrices on $\mathbb H$ and $Q=\{\vert 0\rangle \langle 0\vert , \vert 1\rangle \langle 1\vert \}.$ Then, it is clear that $Z$ is $Q$-classical and $X$ and $Y$ are $Q$-maximally non-classical.

\end{example}

\begin{lemma}
\label{lemNEW4.6_3} Let ${\mathbb{H}}_{1}, {\mathbb{H}}_{2} $ be two separable Hilbert spaces, and  ${{\mathbb{H}}_{1} }{\otimes} {{\mathbb{H}}_{2}}$ be their topological tensor product. Assume that $ \{q_{1} , q_{2},... q_{n}, ...\}$ and $ \{p_{1} , p_{2},... p_{n}, ...\}$ are orthonormal bases of ${\mathbb{H}}_{1}$ and ${\mathbb{H}}_{2}$. Then $\{q_{i}{\otimes}q_{j}, i,j\in{\mathbb{N}}\}$ is an orthonormal basis of ${{\mathbb{H}}_{1}} {\otimes} {{\mathbb{H}}_{2}}. $
\end{lemma}

\noindent{\bf{Proof}.}\quad
The proof is straightforward.
$\hfill\blacksquare$
\bigskip

Let ${\mathbb H}$ and ${\mathbb H}^{\prime}$ be Hilbert spaces. Let $P=\{p_{1}, p_{2}, ..., \}$ and $Q=\{q_{1}, q_{2}, ..., \}$ be complete sets of mutually orthogonal projections of the Hilbert spaces ${\mathbb H}$ and ${\mathbb H}^{\prime}$. Then:
$${P\otimes{Q}}=\{{p_{i}\otimes{q_{j}}} , i,j\in{\mathbb{N}} \}$$
is a complete set of mutually orthogonal projections on ${\mathbb{H}}\otimes{{\mathbb {H}}^{\prime}}.$ Let $T$ (resp. $S$) be a bounded operator on ${\mathbb{H}}$ (resp. ${\mathbb{H}}^{\prime}$). Then:

${T_{P}}\otimes{S_{Q}}=\sum_{n,m}{{{p_{n}}{T}{p_{n}}}\otimes {q_{m}{S}q_{m}}}$
$=\sum_{n,m}{({p_{n}}\otimes{ q_{m}})(T\otimes {S})({p_{n}}\otimes {q_{m}})}$
$=(T\otimes {S})_{P\otimes{Q}}$.
\begin{convention}
\label{con4.1} Let ${\mathbb H}_{1}$ and ${{\mathbb H}_{2}}$ be Hilbert spaces, $T\in B({\mathbb H}_1\otimes{{\mathbb H}}_2)$ and ${T_1}\in B_{+}({\mathbb H}_1)$. We set

 $${T_1}{\color{black}{ \bullet}} T:=({{T_1}^{\frac{1}{2}}}{\otimes}{I_{2}})T ({{T_1}^{\frac{1}{2}}}{\otimes}{I_2})$$
Here $I_2$ is the identity mapping of ${\mathbb H}_2$.

\end{convention}

\begin{lemma}
\label{lemNEW4.6_6} Let ${\mathbb{H}}_{1}$ and ${\mathbb{H}}_{2}$ be separable Hilbert spaces. Let $Q\in{{\pi}_{0}({\mathbb H}_{1})}$ and $P\in{{\pi}_{0}({\mathbb H}_{2})}$. Assume that
$${\dot{\rho}}={\sum}_{i,j=1}^{\infty}{{{\lambda}_{ij}}
{q_{i}{\otimes}{p_{j}}}}\in B_{+}{({{\mathbb{H}}_{1}}{\otimes}{{\mathbb{H}}_{2}})}$$
is such that for all $i \in {\mathbb N}$ , $Tr_{1}(q_{i}{\bullet}{\dot{\rho}})$ is an invertible density matrix on
${\mathbb H}_{2}$. Moreover, assume that
$${\rho}_{1}={\sum}_{i=1}^{\infty}{{\lambda}_{i}q_{i}}$$
is an invertible density matrix on ${\mathbb{H}}_{1}$. Let $ \rho={\rho}_{1}{\bullet}{\dot{\rho}}$. Then

1) ${\rho}_{1}=Tr_{2}(\rho).$

2) $\rho$ is an invertible density matrix on $ {{\mathbb H}_{1}} {\otimes}{{\mathbb H}_{2}}.$

Conversely, assume that $\rho$ is an invertible density matrix on $ {{\mathbb {H}}_{1}} {\otimes}{{\mathbb{H}}_{2}}.$ Then

3) ${\rho}_{1}=Tr_{2}(\rho)$ is an invertible density matrix on ${\mathbb H}_{1}.$

4) Let ${\dot{\rho}}=({\rho}_{1})^{-1}{\bullet}{\rho}$. Then, ${\dot{\rho}}$ is a positive operator on
${{\mathbb H}_{1}}{\otimes}{{\mathbb H}_{2}},$ and for each $i\in{\mathbb N}$, $Tr_{1}(q_{i}{\bullet}{\dot{\rho}})$ is an invertible density matrix on ${\mathbb H}_{2}.$
\end{lemma}

\noindent{\bf{Proof}.}\quad

1) $$\rho={\rho}_{1}{\bullet}{\dot{\rho}}=({\sum}_{i=1}^{\infty}{{\lambda}_{i}q_{i}})^{1/2}{\otimes}{I_{2}}(
{\sum}_{i,j=1}^{\infty}{{{\lambda}_{ij}}{q_{i}{\otimes}{p_{j}}}}) ({\sum}_{i=1}^{\infty}{{\lambda}_{i}q_{i}})^{1/2}{\otimes}{I_{2}}=$$
$$({\sum}_{i=1}^{\infty}{{\lambda}_{i}^{1/2}{q_{i}}}){\otimes}{I_{2}}( {\sum}_{i,j=1 }^{\infty}{{{\lambda}_{ij}}{q_{i}{\otimes}{p_{j}}}}) ({\sum}_{i=1}^{\infty}{{\lambda}_{i}^{1/2}{q_{i}}}){\otimes}{I_{2}}=$$
$${\sum}_{i,j=1}^{\infty}{{\lambda}_{i}{\lambda}_{i,j}q_{i}{\otimes}p_{j}}=
{\sum}_{i=1}^{\infty}{{\lambda}_{i}q_{i}{\otimes}{\sum}_{j=1}^{\infty}{\lambda}_{ij}{p_{j}}}.$$
Therefore, $$Tr_{2}(\rho)=Tr_{2}( {\sum}_{i=1}^{\infty}{{\lambda}_{i}q_{i}{\otimes} {\sum}_{j=1}^{\infty}{\lambda}_{i,j} p_{j}})=
{\sum}_{i=1}^{\infty}{{\lambda}_{i}({\sum}_{j=1}^{\infty}{\lambda}_{ij})q_{i}}.$$
But $${\sum}_{j=1}^{\infty}{\lambda}_{i,j}=Tr(Tr_{1}(q_{i}{\bullet}{\dot{\rho}}))=1.$$
Therefore, ${\rho}_{1}=Tr_{2}(\rho).$

2) Since for each $i\in{\mathbb N}$,
$${\sum}_{j=1}^{\infty}{{\lambda}_{ij}{p_{j}}}= Tr_{1}(q_{i}{\bullet}{\dot{\rho}})$$
and ${\rho}_{1}={\sum}_{i=1}^{\infty}{{\lambda}_{i}q_{i}}$ are positive and invertible, for all $i$, $j\in{\mathbb N}$,
${\lambda}_{i}{\lambda}_{ij}>{0}.$ Therefore, $\rho$ is invertible.

3) Let $\rho= {\sum}_{i,j=1}^{\infty}{{{\mu}_{ij}}{q_{i}{\otimes}{p_{j}}}}$. Then $${\rho}_{1}=Tr_{2}(\rho)=Tr_{2}({\sum}_{i,j=1}^{\infty}{{{\mu}_{ij}}{q_{i}{\otimes}{p_{j}}}})
= {\sum}_{i=1}^{\infty}{({\sum}_{j=1}^{\infty}{\mu}_{ij})q_{i}}.$$
Since $\rho$ is positive and invertible, for all $i$, $j\in {\mathbb N}$ , $\mu_{ij}>{0}.$ Moreover, $Tr(\rho_{1})=Tr(\rho)=1.$ Therefore, $\rho_{1}=Tr_{2}(\rho)$ is an invertible density matrix on
${\mathbb H}_{1}.$

4) It is evident that ${\dot{\rho}}$ is a positive operator on ${{\mathbb H}_{1}}{\otimes}{{\mathbb H}_{2}}.$
Clearly, ${\rho}_{1}^{-1}=(Tr_{2}(\rho))^{-1}= {\sum}_{i=1}^{\infty}{({\sum}_{j=1}^{\infty}{\mu}_{ij})^{-1}q_{i}}.$ Therefore,
${\dot{\rho}}={\sum}_{i,j=1}^{\infty}{({\sum}_{k=1}^{\infty}{\mu}_{ik})^{-1}{\mu}_{ij}q_{i}{\otimes}
p_{j}}$, and for $i\in{\mathbb{N}}$, $Tr_{1}(q_{i}{\bullet}{\dot{\rho}})= {\sum}_{j=1}^{\infty}{({\sum}_{k=1}^{\infty}{\mu}_{ik})^{-1}{\mu}_{ij}
p_{j}}$. Clearly $Tr_{1}(q_{i}{\bullet}{\dot{\rho}})$ is a positive operator and $Tr(Tr_{1}(q_{i}{\bullet}{\dot{\rho}}))=1.$ Therefore, for each $i\in {\mathbb N}$ , $Tr_{1}(q_{i}{\bullet}{\dot{\rho}})$ is an invertible density matrix on ${\mathbb H}_{2}.$

$\hfill\blacksquare$
\bigskip

\begin{corollary}

 1) If ${\rho}$ is not invertible but for some $i\in{\mathbb{N}}$ , $Tr_{1}{(q_{i}{\bullet}{\rho})}$ is not zero, ${\rho}_{1}$ is not zero at $q_{i}.$ In this case, ${\dot{\rho}}={{\rho}_{1}^{-1}}{\bullet}{\rho}$ is well defined at all points $ q_{i}{\otimes}{p_{j}} , j\in{\mathbb{N}}.$

2) ${\rho}_{1}$ is zero at $q_{i}$ if and only if $\rho$ is zero at all points $q_{i}{\otimes}{p_{j}}    , j\in{\mathbb{N}}.$ In this case ${\dot{\rho}}={{\rho}_{1}^{-1}}{\bullet}{\rho}$ is indeterminate at all
$ q_{i}{\otimes}{p_{j}} , j\in{\mathbb{N}}.$

\end{corollary}

\begin{lemma}
\label{lemNEW4.6_1} Let ${\mathbb{H}}_{1}$ and ${\mathbb{H}}_{2}$ be separable Hilbert spaces. Let $Q\in{{\pi}_{0}({\mathbb H}_{1})}$ and $P\in{{\pi}_{0}({\mathbb H}_{2})}$. Assume that
$${\rho}\in D{({{\mathbb{H}}_{1}}{\otimes}{{\mathbb{H}}_{2}})}.$$
Then, $(Tr_{2}(\rho))_{Q}= Tr_{2}({\rho}_{(Q{\otimes}P)}).$
\end{lemma}

\noindent{\bf{Proof}.}\quad Clearly, in genral $\rho$ can be written as follows
$$\rho={\sum}_{i,j=1}^{\infty}{{\lambda}_{ij}|ij><ij|},$$
where, ${\bf{{b}_{1}}}=\{|i> i\in {\mathbb N} \}$ and ${\bf{{b}_{2}}}=\{|j> j\in {\mathbb N} \}$ are orthonormal bases of ${\mathbb{H}}_{1}$ and ${\mathbb{H}}_{2}$. Then,
$$(Tr_{2}(\rho))_{Q}={\sum}_{i=1}^{\infty}{q_{i}(Tr_{2}(\rho))q_{i}}=
{\sum}_{i=1}^{\infty}{q_{i}(Tr_{2}({\sum}_{k,l=1}^{\infty}{{\lambda}_{kl}|kl><kl|}) )q_{i}}=$$
$${\sum}_{i=1}^{\infty}{{\sum}_{k=1}^{\infty}{({\sum}_{l=1}^{\infty}{{\lambda}_{kl}}q_{i}|k><k|}q_{i})}.$$
On the other hand, $$Tr_{2}({\rho}_{(Q{\otimes}P)})=Tr_{2}({\sum}_{i,j=1}^{\infty}{{\sum}_{k,l=1}^{\infty}
{{\lambda}_{kl}q_{i}|k><k|q_{i}{\otimes}p_{j}|l><l|p_{j}}})=$$
$$Tr_{2}({\sum}_{i=1}^{\infty}{{\sum}_{k=1}^{\infty}{{\sum}_{l=1}^{\infty}
{{\lambda}_{kl}q_{i}|k><k|q_{i}{\otimes}|l><l|{\sum}_{j=1}^{\infty}{p_{j}}}}})=$$
$$ Tr_{2}({\sum}_{i=1}^{\infty}{{\sum}_{k=1}^{\infty}{{\sum}_{l=1}^{\infty}
{{\lambda}_{kl}q_{i}|k><k|q_{i}{\otimes}|l><l|}}})=$$
$${\sum}_{i=1}^{\infty}{{\sum}_{k=1}^{\infty}{{\sum}_{l=1}^{\infty}
{{\lambda}_{kl}q_{i}|k><k|q_{i}}}}.$$
Therefore, $(Tr_{2}(\rho))_{Q}= Tr_{2}({\rho}_{(Q{\otimes}P)}).$

$\hfill\blacksquare$
\bigskip

\section{Quantum Model,Quantum Source and Quantum Strategy}

As we said in the introduction our work in this paper inspired by the
Minimum Description Length Principle
is based on universal quantum sources associated with quantum models. In this part{\color{black}{,}} we define several versions of universal quantum sources associated with a quantum model and investigate some of their properties. In the same section{\color{black}{,}} we prove the existence of universal quantum sources and give a constructive way to build them. We also define quantum strategy\index{quantum strategy} and treat its relation to universal quantum source{\color{black}{s}}.

Before going further in this section let us give some comments on the use of semi-density matrices and on our definition of universal quantum sources.

The minimum description length principle is a powerful tool in statistical (inductive) inference. It is essentially based on two important notions:

\subsection*{2-part coding}
The estimation by 2-part code\index{2-part code} can be considered as a mathematical formulation of Occam's Razer\index{Occam's Razer} which says that between different descriptions of a data set, the simpler is the better. Assume that these descriptions are encoded in such a way that {\color{black}{they}} reflect their complexities. Then the description with the shortest code-length is the better.

More precisely, let $\cal M$ be a nonempty set of probability density (mass) functions on a set $\cal X$ and let $D\subset{\cal X}^{n}$ be an i.i.d data set generated by $p^{\star}\in {\cal M}$. Assume that elements of $\cal M$ are encoded. For each $p\in {\cal M}${\color{black}{,}} the length of its associated code-word will be denoted by $L(p)$ and $-log_{2}{p(D)}$ will be denoted by $L(D|p).$ Let

$$\ddot{p}=argmin_{p\in{\cal{M}}}{(L(p)+L(D|p))}.$$

Clearly for each $p\in {\cal M}$, $L(p)+L(D|p)$ is the length of an encoded description of the data set $D$ and
$ \ddot{p}$ is chosen according to Occam's Razer.

\subsection*{Universal coding\index{Universal coding}}
Under above assumptions on $\cal M$ and $\cal X$, assume that for each $n\in {\mathbb N}$, ${\bar{p}}^{(n)}$is a probability density (mass) function on ${\cal X}^{n}$. The sequence
$\bar{p}=({\bar{p}}^{(n)})_{n\in {\mathbb N}}$  of probability density (mass) functions will be called universal with respect to $\cal M$, if for each ${\epsilon}>0$, each $p\in {\cal M}$, there exists
$n_{0}\in{\mathbb N}$ such that for all $ n\geq{n_{0}}$ and all $x^{(n)}\in{{\cal X}^{n}}$ we have

$$-log_{2}{{\bar{p}}^{(n)}{(x^{(n)})}}-(-log_{2}{{p}^{(n)}{(x^{(n)})}})\leq{n{\epsilon}}.$$
For more details see [4].

Now let us explain briefly the way we have gone through to quantize these two notions.

Let the Hilbert space $\mathbb H$ be the state space of a quantum system $A$, which is prepared in an unknown state ${\rho}_{0}${\color{black}{,}} a density matrix on $\mathbb H$, and let $Q=\{q_{m} | m\in {O} \}{\color{black}{\in \pi_0(\mathbb{H})}}$ where $O$ is the set of outcomes{\color{black}{,}} be a projective quantum measurement system. Assume that
$\cal{M}$ is a nonempty set of density matrices on $\mathbb{H}$ and $D\in{O^{n}}$ is the set of outcomes of performing {\color{black}{the }}$Q$-measurement on $n$ quantum systems identical to $A$ and prepared in the same state ${\rho}_{0}$.
In performing {\color{black}{the }}$Q$-measurement on the quantum system $A$ in an arbitrary state $\rho$ the probability of outcome $m$ is
$$\mathbb{P}(m)=Tr(q_{m}{\rho}) =Tr( q_{m}{\rho}q_{m})$$

\subsection*{ 2-part coding ${\longrightarrow} $ {semi-density matrix}}

Let elements of $\cal M$ be somehow encoded and for each ${\rho}\in{ \cal M}$ let $L(\rho) $be the length of the code-word associated with $\rho$ and let $L(D|{\rho})=-log_{2}{Tr({\otimes}^{m\in {D }}{{q_{m}{\rho}q_{m}}})}.$ Then for each ${\rho}\in{ \cal M}$ we have
$$ L(\rho)+ L(D|\rho)=-log_{2}{2^{- L(\rho)}}- log_{2}{Tr({\otimes}^{m\in { D}}{q_{m}{\rho}q_{m}})}$$
$$=-log_{2}{({2^{- L(\rho)}} {Tr({\otimes}^{m\in {D}}{q_{m}{\rho}q_{m}})})}
=-log_{2}{Tr(2^{- L(\rho)}{{\otimes}^{m\in {D}}{q_{m}{\rho}q_{m}}})}$$

$$=-log_{2}{Tr({{\otimes}^{m\in {D}}{q_{m}}} {(2^{-L(\rho)}{\rho}^{(n)})} {{\otimes}^{m\in {D }}{q_{m}}})}.$$
But the function $ log_{2}$ is increasing and $Tr( {{\otimes}^{m\in {D}}{q_{m}}} {(2^{-L(\rho)}{\rho}^{(n)})} {{\otimes}^{m\in {D}}{q_{m}}})$ is also increasing with respect to the
semi-density matrices\index{semi-density matrices} $2^{-L(\rho)}{{\rho}^{(n)}}$. As in the above classical case
$$\ddot{\rho}=argmin_{{\rho}\in{\cal{M}}}{L(\rho)+L(D|{\rho})}=argmax_{{\rho}\in{\cal{M}}}
{({\otimes}^{m\in {D}}{q_{m}})} {(2^{-L(\rho)}{{\rho}^{(n)}})} {({\otimes}^{m\in {D}}{q_{m}})}$$
is an estimation of the state of the system ${\rho}_{0}$ according to Occam's Razer. Notice that ${(2^{-L(\rho)}{{\rho}^{(n)}})}$ is a semi-density matrix.

\subsection*{ Universal Coding $\longrightarrow$ { Universal {\color{black}{ Density Matrix}}}}

Let ${\bar{\rho}}^{(n)}$ and ${\bar{{\rho}^{\prime}}}^{(n)}$ be two density matrices on ${\mathbb{H}}^{(n)}.$ Assume that as in classical case for ${\epsilon}>{0}$ there exists $n_{0}\in {\mathbb{N}}$
such that for all $n>{n_{0}}$ and for all $q^{(n)}\in {Q^{(n)}}$ we have
$$-{ log_{2}{Tr(q^{(n)}{{\bar{\rho}}^{(n)}}{q^{(n)}})}}-{(-log_{2}{ Tr(q^{(n)}{ {\bar{{\rho}^{\prime}}}^{(n)}} {q^{(n)}})})}\leq{n{\epsilon}}.$$
From the above inequality we have

${ log_{2}{Tr(q^{(n)}{{\bar{\rho}}^{(n)}}{q^{(n)}})}}\geq{log_{2}{2^{-n{\epsilon}}}+log_{2}{ Tr(q^{(n)}{ {\bar{{\rho}^{\prime}}}^{(n)}} {q^{(n)}})}}$$
$$= log_{2}{Tr(q^{(n)}{({2^{-n{\epsilon}}} {\bar{{\rho}^{\prime}}}^{(n)}}) {q^{(n)}})}$
But the inequality
$$ log_{2}{Tr(q^{(n)}{{\bar{\rho}}^{(n)}}{q^{(n)}})}\geq{ log_{2}{Tr(q^{(n)}{({2^{-n{\epsilon}}} {\bar{{\rho}^{\prime}}}^{(n)}}) {q^{(n)}})}}$$
is equivalent to
$$q^{(n)}({\bar{\rho}}^{(n)}-{2^{-n{\epsilon}}{\bar{{\rho}^{\prime}}}^{(n)}})q^{(n)}\geq{0}.$$

In the following all tensor products of Hilbert spaces are topological tensor products.

The $n$-times tensor product of{\color{black}{ a}} Hilbert space $\mathbb H$  will be denoted by ${\mathbb H}^{(n)}$ and in general, for each $T\in B(\mathbb{H})$, ${T}^{{\otimes{n}}}:={\bigotimes}^n {T}$. The sequence
$({\mathbb{H}}^{(n)})_{n\in {\mathbb{N}}}$ of Hilbert spaces will be denoted by ${\mathbb{H}}^{\star}$ and for
$T_{(n)}\in {\mathbb{H}}^{(n)}$ the sequence $(T_{(n)})_{n\in {\mathbb{N}}}$ will be denoted by $ {T^{\otimes}}.$ In this case we say that ${T^{\otimes}}$ is an operator on ${\mathbb{H}}^{\star}$ and if for all $n\in {\mathbb{N}}$, $ T_{(n)}$ is a (semi-)density matrix, then  $ {T^{\otimes}}$ will be called a (semi-)density matrix on ${\mathbb{H}}^{\star}$. A semi-density matrix $T^{\otimes}= (T_{(n)})_{n\in {\mathbb{N}}}$ on ${\mathbb{H}}^{\star}$ is called nonzero if for all $n\in {\mathbb{N}}$, $ T_{(n)}\neq {0}$. In this case the associated density matrix of $T^{\otimes}$ is $\omega{(T^{\otimes})}=(\frac{T_{(n)}}{Tr(T_{(n)})})_{n\in {\mathbb{N}}}$. From now on semi-density matrices on ${\mathbb{H}}^{\star}$ will be denoted by $\bar{\rho}=(\bar{\rho}^{(n)})_{n\in {\mathbb{N}}}$. The semi-density matrix $\bar{\rho}=(\bar{\rho}^{(n)})_{n\in {\mathbb{N}}}$ will be called
\begin{enumerate}
\item simple if ${\bar{\rho}}^{(1)}={\rho}$ and for each $n\in{\mathbb{N}}$, ${\bar{\rho}}^{(n)}$ is the tensor product of $\rho$ and ($n-1$)-times tensor product of ${\omega}(\rho).$

\item a generalized quantum source if for each $1<n\in{\mathbb N}$, ${\bar{\rho}}^{(n-1)}=Tr_{n}({\bar{\rho}}^{(n)})$.

\item regular if for each $n$, ${\bar{\rho}}^{(n)}$ is invertible

\end{enumerate}
When for each $n\in{\mathbb N}$, $Tr({\bar{\rho}}^{(n)})=1$, the generalized quantum source
$ \bar{\rho}=( {\bar{\rho}}^{(n)})_{n\in {\mathbb N}}$ will be called a quantum source.

Let $Q=\{q_{1} , q_{2} , ... , q_{n} , ... \}{\color{black}{\in \pi(\mathbb{H})}}$ be a complete set of mutually orthogonal projections of the Hilbert space $ {\mathbb H}$ and let $I=(i_{1} , i_{2} , ... , i_{n})\in {\mathbb{N}}^{(n)}$. Then the projection
$q_{i_{1}}\bigotimes{q_{i_{2}}}\bigotimes... \bigotimes{q_{i_{n}}}$ will be denoted by
$q_{I}^{(n)}$ or simply by $q^{(n)}$ if there is no ambiguity. The set $\{q_{I}^{(n)} | I\in {\mathbb{N}}^{(n)}\}$ will be denoted by $Q^{(n)}.$

\begin{definition}
Let $\mathbb{H}$ be a separable Hilbert space. Each non-empty set $\mathcal{M}$ of (semi-)density matrices on $\mathbb{H}$ will be called a (generalized) quantum model.
\end{definition}

{\color{black}{
\begin{definition}
Let M be a generalized quantum model and $( { M} , {\Sigma}, {\mu} )$ be a measure space. Then M  will be called \textbf{Bayesian} if $\int_M{\rho}d{\mu}(\rho)$ exists and is a density matrix. From now on, when there is no ambiguity  the triple 
$(M , {\Sigma} , {\mu})$ will be denoted by $M$.
\end{definition}
}}  
\begin{lemma}
\label{lem5.1} Let $\underline{\mathcal M}$ be a Bayesian generalized quantum model which is a measure space and
let $\bar{\rho}^{(n)}=\int_{\underline{\mathcal M}}{\rho}^{(n)}d\mu(\rho).$ Then, the sequence $({\bar{\rho}}^{(n)})_{n\in{\mathbb N}}$ is a quantum source. Which is called the quantum source associated with $\underline{\mathcal M}$.
\end{lemma}

\noindent{\bf{Proof}.}\quad For each $n\in{\mathbb N}$ clearly we have $Tr_{n+1}({\rho}^{(n+1)})={\rho}^{(n)}.$ Therefore,

$$Tr_{n+1}({\bar{\rho}}^{(n+1)})=\int_{\underline{\cal M}}{Tr}_{n+1}({\rho}^{(n+1)})d\mu(\rho) =
\int_{\underline{\cal M}}{\rho}^{(n)}d\mu(\rho)={\bar{\rho}}^{(n)}.$$

$\hfill\blacksquare$

\begin{lemma}
\label{lem5.5} Let $U\in {B(\mathbb H)}$ be a unitary operator and ${\bar {\rho}}$ be a quantum source. Then $U{\bar {\rho}}{U^{\dagger}}=(U^{(n)}{\bar{\rho}}^{(n)}(U^{\dagger})^{(n)})_{n\in {\mathbb N}}$ is also a quantum source.
\end{lemma}

\noindent\textbf{Proof.}\quad

Obviously any element $\bar{\rho}^{(n+1)}\in B(\mathbb{H}^{(n+1)})$ can be written as

${\bar {\rho}}^{(n+1)}={\sum}_{i,j}R_{i,j}\otimes\vert i \rangle\langle j\vert$ where $R_{i,j}\in B(\mathbb{H}^{(n)})$. Because $\bar{\rho}$ is a quantum source we have
$${Tr}_{n+1}({\bar{{\rho}}^{(n+1)}})={\sum}_{i}{R_{i,i}}={\bar{{\rho}}^{(n)}}$$
So,
\begin{equation}
\begin{array}{rl}
(U{\bar{\rho}}{U^{\dagger}})^{(n+1)}&=U^{(n+1)}{\bar{\rho}}^{(n+1)}{( U^{\dagger})}^{(n+1)} \\
&={\sum}_{i,j=1}^{\infty}(U^{(n)}{R_{i,j}}{(U^{\dagger})}^{(n)}){\otimes}U{\vert i \rangle\langle j\vert}U^{\dagger}.
\end{array}
\end{equation}

Therefore,

\begin{equation}
\begin{array}{rl}
{Tr}_{n+1}(U{\bar{\rho}}{U^{\dagger}})^{(n+1)}&={\sum}_{i,j=1}^{\infty}(U^{(n)}{R_{i,j}}{(U^{\dagger})}^{(n)})Tr(U{\vert i \rangle\langle j\vert}U^{\dagger}) \\
&={\sum}_{i=1}^{\infty}(U^{(n)}{R_{i,i}}{(U^{\dagger})}^{(n)})\\
&=U^{(n)}({\sum}_{i=1}^{\infty}{R_{i,i}}){(U^{\dagger})}^{(n)}\\
&=U^{(n)}{\bar{\rho}}^{(n)}(U^{\dagger})^{(n)}=(U{\bar{\rho}}{U^{\dagger}})^{(n)}
\end{array}
\end{equation}

Therefore, $U{\bar{\rho}}U^{\dagger}$ is a quantum source.

$\hfill\blacksquare$

{{\color{black}{
In this work $ln$ denotes natural logarithm and $log$ denotes logarithm in base $2$.
\begin{definition}
Let $\rho$ and ${\rho}^{\prime}$ be density matrices. Then the quantum relative entropy of $\rho$ and ${\rho}^{\prime}$ is
$$S({\rho}\Vert {\rho}^{\prime})=tr({\rho}log{\rho})-tr({\rho}log{\rho}^{\prime})$$
\end{definition}
}}}
\begin{definition}
\label{def5.5} Let ${\mathcal M}$ be a
quantum model and $\bar{\rho}=({\bar{\rho}}^{(n)})_{n\in {\mathbb N}}$ be a semi-density matrix on ${\mathbb H}^{\star}$. Let $Q\in{{\pi}(\mathbb H)}$. We say that $\bar{\rho}$  is

\begin{enumerate}
\item \textit{Universal\index{Universal}} relative to ${\cal{M}}$ if for each ${\rho}\in {{\cal {M}}}$ and for each $\epsilon>{0}$ there exists an $n_{0}\in {\mathbb{N}}$ such that for all $n\geq n_{0}$ we have:
$${\bar{\rho}}^{(n)}-2^{-n{\epsilon}}{\rho}^{(n)}\geq{0}.$$
\item \textit{Universal in the expected sense\index{Universal in the expected sense}}{\color{black}{ relative}}  to ${\mathcal M}$ if:
$$S({{\rho}}^{(n)}\Vert{\bar{\rho}}^{(n)})\leq{n{\epsilon}}.$$
\item \textit{$Q$-Universal relative\index{$Q$-Universal}} to ${\cal{M}}$ if for each ${\rho}\in {{\cal{M}}}$ and for each $\epsilon>{0}$ there exists an $n_{0}\in {\mathbb{N}}$ such that for all $n\geq n_{0}$ we have:
$${\bar{\rho}}_{Q}^{(n)}-2^{-n{\epsilon}}{\rho}_{Q}^{(n)}\geq{0}.$$

\item \textit{$Q$-universal relative } to ${\cal M}$ in the expected sense if $$S({{\rho}_{Q}}^{(n)}\Vert{\bar{\rho}_{Q}}^{(n)})\leq{n{\epsilon}}.$$

\item \textit{weakly universal relative} to ${\cal M}$ if for each $Q\in\pi_0(\mathbb{H})$ $\bar\rho_{Q}$ is {$Q$-universal relative } to ${\cal M}$ .

\end{enumerate}

In the above if $\epsilon$ does not depend on ${\rho}$, $\bar{\rho}$ is called uniformly (Q-)universal.

\end{definition}

\begin{lemma}
\label{lem5.6} With the above notations and conventions{\color{black}{,}} 1 implies 2 and 3.
\end{lemma}
{\bf{Proof.}$1{\longrightarrow}{2}$}

Clearly we have

\begin{equation}
\begin{array}{rl}
&{\bar{\rho}}^{(n)}-2^{-n{\epsilon}}{\rho}^{(n)}\geq{0}\\
\Rightarrow& n{\epsilon}+{log{\bar{\rho}}^{(n)}}-{log{\rho}}^{(n)}\geq{0}\\
\Rightarrow& n{\epsilon}{\rho}^{(n)}+{({\rho}^{(n)})^{1/2}}({{log{\bar{\rho}}^{(n)}}-{log{\rho}}^{(n)}}){({\rho}^{(n)})^{1/2}}\geq{0}\\
\Rightarrow& Tr(n{\epsilon}{\rho}^{(n)}+{({\rho}^{(n)})^{1/2}}({{log{\bar{\rho}}^{(n)}}-{log{\rho}}^{(n)}}){({\rho}^{(n)})^{1/2}})\geq{0}\\
\Rightarrow& n{\epsilon}+{\rm Tr}{\rho}^{(n)}{(log{\bar{\rho}^{(n)}}-log{\rho}^{(n)})}\geq{0}.\\
\Rightarrow& S({{\rho}}^{(n)}\Vert {\bar{\rho}}^{(n)})\leq{n{\epsilon}}.
\end{array}
\end{equation}
The other part is clear.
$\hfill\blacksquare$
\begin{example}
\label{ex5.1} Let $\underline{\cal M}$ be a Bayesian countable generalized quantum model consisting of nonzero semi-density matrices and let ${\mathcal M}$ be its associated quantum model. Then for each element ${{\rho}^{\ast}}\in{\underline{\mathcal M}}$ and each $n\in{\mathbb N}$ we have

$${{\bar{\rho}}^{(n)}}={\sum}_{{\rho}\in{\underline{\cal M}}}{{\rho}^{(n)}}\geq{{\rho}^{\ast}}^{(n)}.$$

Now let $\epsilon$ be given and let ${n_{0}}\in {\mathbb N}$ be such that

$${\rm Tr}({\rho}^{\ast})\geq{2^{-{(n_{0})}{\epsilon}}}.$$

Then, for each $n\geq{n_{0}}$ we have

$${\bar{\rho}}^{(n)}-2^{-n{\epsilon}}{{\rho}}^{(n)}\geq{0},$$

where ${\rho}={\omega}( {{\rho}^{\ast}})$. Therefore, $\bar{\rho}=({\bar{\rho}}^{(n)})_{n\in{\mathbb{N}}}$ is universal{\color{black}{ relative to}} ${\mathcal M}.$

\end{example}

\begin{example}
\label{ex6.2} Let ${\cal M}$ be a quantum model and let ${\bar{\rho}}$ be a universal density matrix{\color{black}{ relative to}} ${\cal M}$ and $U$ be a unitary operator. Then $U{\bar{\rho}}U^{-1}$ is a universal density matrix{\color{black}{ relative to}} $U{\cal M} U^{-1}$ where $U{\cal M} U^{-1}=\{U\rho U^{-1}\vert \rho \in{\cal M}\} $
\end{example}

Let $\mathbb H$ be an $m-$dimensional Hilbert space and $Q=\{q_{1} , q_{2} ,  ... ,   q_{m} \}     \in{\pi}_{0}(\mathbb H)$. Let $\cal M$ be a quantum model which is a compact Riemannian submanifold of $B_{T}{(\mathbb H)}$ consisting of regular density matrices. Assume that ${\epsilon}>{0}$ is given. Let ${\rho}\in{\cal{M}}$ be arbitrary. Let ${c_{\rho}}={min}_{q\in{Q}}{Tr(q{\rho}q)}$ and ${{\delta}_{\rho}}={c_{\rho}}(1-{2}^{-{\epsilon}/2})$. For each $1\leq{i}\leq{m}$, let
${\rho}_{i}=argmax_{{\rho}^{\prime}\in{B(\rho , {{\delta}_{\rho}}/2)}}{q_{i}{\rho}^{\prime}{q_{i}}}$. Then, for each ${{\rho}^{\prime}}\in{B(\rho , {{\delta}_{\rho}}/2)}$ we have

$${\rm Tr}(q_{i}{{\rho}_{i}}q_{i})-{\rm Tr}{({q_{i}{{\rho}^{\prime}}q_{i})}=
{{\rm Tr}(q_{i}{{\rho}_{i}}q_{i}-{q_{i}{{\rho}^{\prime}}q_{i})}}={\Vert q_{i}({{\rho}_{i}}-{{\rho}^{\prime}})q_{i}\Vert }}$$
$$\leq{\Vert {{\rho}_{i}}-{{\rho}^{\prime}}\Vert }\leq{d({\rho}_{i} , {{\rho}^{\prime}})} \leq{{\delta}_{\rho}}.$$

and it is straightforward to see that

$$(\star) ,  q_{i}{{\rho}^{\prime}}q_{i}\geq{2^{-{\epsilon}/2}q_{i}{{\rho}_{i}}q_{i}}.$$

Now, let $ {\beta}:{\cal{M}}\longrightarrow{[0 , \infty[}$ be an almost nonzero continuous function with ${\int}_{\cal{M}}{\beta({\rho}^{\prime})}\leq{1}.$ Since $\cal{M}$ is compact there exists a finite number of $ B({\rho}, {{\delta}_{\rho}}/2)$'s that cover $\cal{M}$. The set consisting of centers of this finite sets will be denoted by $\Sigma$ and the nonzero real number $min_{{\rho}\in{\Sigma}}{{\int}_{ B({\rho}, {{\delta}_{\rho}}/2)}{{\beta}({\rho}^{\prime})}dvol_{\cal{M}}({\rho}^{\prime})} $ by ${\Omega}$. Let $k\in{\mathbb{N}}$ be such that ${\Omega}\geq{2^{-k{\epsilon}/2}}.$

Finally, for each $n\in{\mathbb N}$ let ${\bar{\rho}}^{(n)}={\int}_{\cal{M}}{{\beta} ({\rho}^{\prime}){{\rho}^{\prime}}^{(n)}}dvol_{\cal{M}}{({\rho}^{\prime})}$.

Now under the above notations and conventions, we have the following theorem.

\begin{theorem}
\label{theo5.1} The quantum source  $({\bar{{\rho}}}^{(n)})_{n\in{\mathbb{N}}}$ is uniformly weakly universal for $\cal{M}$.
\end{theorem}

{\bf{Proof}} Since the set $\{B(\rho , {{\delta}_{\rho}}/2) , {\rho}\in{\Sigma}\}$ covers $\cal{M}$, each ${\rho}^{\prime}\in{\cal{M}}$ is in some $B(\rho , {{\delta}_{\rho}}/2).$ Let us denote the set
$\{i\in{\mathbb{N}} | 1\leq{i}\leq{m} \}$ by $[1 , m].$ Let $n\in{\mathbb{N}}$ be arbitrary and
$I=(i_{1} , i_{2} , ... , i_{n})\in{[1 , m]^{n}}.$ Assume that
${\rho}\in{\Sigma}$  and   ${\rho}^{\star}\in{B(\rho , {{\delta}_{\rho}}/2) }$. Then, we have

$${q_{I}^{(n)}{\bar{\rho}}^{(n)}q_{I}^{(n)}}={\int}_{\cal{M}}{{\beta}({\rho}^{\prime})} {q_{I}^{(n)}{{\rho}^{\prime}}^{(n)}
q_{I}^{(n)}{d{vol}_{\cal{M}}{({\rho}^{\prime})}}}\geq{\int}_{B(\rho , {{\delta}_{\rho}}/2)} { {\beta}({\rho}^{\prime}) {q_{I}^{(n)}{{\rho}^{\prime}}^{(n)}q_{I}^{(n)}{d{vol}_{\cal{M}}{({\rho}^{\prime})}}}}$$
$$\geq({2^{-n{\epsilon}/2}}{{\Omega}}){\otimes}_{l=1}^{n}{q_{i_{l}}{{\rho}_{i_{l}}}q_{i_{l}}}\geq
2^{-(k+n){\epsilon}/2}q_{I}^{(n)}{{\rho}^{\star}}^{(n)}q_{I}^{(n)}.$$

Now assume that$k=n_{0}.$ Then for each ${\rho}^{\star}\in{\cal{M}}$ and each $n\geq{n_{0}}$ we have

$${q_{I}^{(n)}{\bar{\rho}}^{(n)}q_{I}^{(n)}}-2^{-n{\epsilon}} q_{I}^{(n)}{{\rho}^{\star}}^{(n)}q_{I}^{(n)}\geq{0}.$$

From the above inequality it follows that for all ${\rho}^{\star}\in{\cal{M}}$ we have

$${\bar{\rho}}_{Q}^{(n)}={\sum}_{q^{(n)}\in {Q^{(n)}}}({q^{(n)}{{\bar{\rho}}^{(n)}}}{q^{(n)}})\geq{2^{-n{\epsilon}}{\sum}_{ q^{(n)}\in {Q^{(n)}}}({q^{(n)}{{{\rho}^{\star}}^{(n)}}}{q^{(n)}} })=2^{-n{\epsilon}}{{\rho}^{\star}}_{Q}^{(n)}.$$

Since our proof and its consequences do not depend on $Q\in{{\pi}_{0}{(\mathbb{H})}}$, the quantum source ${\bar{\rho}}=({\bar{\rho}}^{(n)})_{n\in{\mathbb N}}$ is uniformly weakly universal for $\cal M.$

$\hfill\blacksquare$
\bigskip

\begin{corollary}
Under the above conventions and notations , let $\cal M$ 
be commutative. Then, $({\bar{\rho}}^{(n)})_{n\in N}$ is a uniformly universal quantum source for $\cal M$.
\end{corollary}

\begin{lemma}
\label{lem5.7} $S_{\mathcal M}$ the set of all universal quantum source{\color{black}{ relative to}} the quantum model $\mathcal M$ is convex.
\end{lemma}

\noindent{\bf{Proof}.}\quad Let $\bar{{\rho}_{1}}$ and $\bar {{\rho}_{2}}$ be two universal quantum source{\color{black}{relative to}} the quantum model $\mathcal{M}$. Let ${\rho}\in \mathcal{M}$ and ${\epsilon}> {0}$ be given. Then there exists ${n_{0}}\in{\mathbb{N}}$ such that for $k=1, 2$ and $n\geq n_0$ we have:
$${\bar{\rho}}_{k}^{(n)}-2^{-n{\epsilon}}{{\rho}}^{(n)}\geq{0}.$$ Let $\alpha$ and $\beta$ be two positive real numbers such that ${\alpha}+{\beta}=1$. Then
$${\alpha}{\bar{\rho}}_{1}^{(n)}+{\beta}{\bar{\rho}}_{2}^{(n)}-2^{-n{\epsilon}}{\rho}^{(n)}=$$
$${\alpha}({\bar{\rho}}_{1}^{(n)}-2^{-n{\epsilon}}{\rho}^{(n)})+{\beta}({\bar{\rho}}_{2}^{(n)}-2^{-n{\epsilon}}{\rho}^{(n)})\geq{0}.$$
Therefore $S_{\mathcal M}$ at each level $n$ is convex. On the other hand,
$$({\alpha}{\bar{\rho}}_{1}+{\beta}{\bar{\rho}}_{2})^{(n)}={\alpha}{\bar{{\rho}}_{1}}^{(n)}+{\beta}{{\bar{\rho}}_{2}}^{(n)}\in ({S_{\mathcal{M}}})^{(n)},$$
where $({S_{\mathcal M}})^{(n)}=\{\bar{\rho}^{(n)}\vert\bar{\rho}\in S_{\mathcal M}\},$
Therefore $${{\alpha}{{\bar{\rho}}_{1}}+{\beta}{{\bar{\rho}}_{2}}}\in {S_{\mathcal{M}}}.$$

$\hfill\blacksquare$

Before going further it is better to introduce the notion of conditional density matrix.

Let ${\mathbb H}_{1}$ and ${\mathbb H}_{2}$ be Hilbert spaces. Let ${\rho}$ be a density matrix on the Hilbert space ${\mathbb H}_{1}\otimes{\mathbb H}_{2},${\color{black}{$\rho_{1}=Tr_{2}(\rho)$ and}} ${\rho}_{2 | 1}={\rho}_{1}^{-1}\bullet{\rho}.$ When
${\mathbb{H}}_{1}={ \mathbb{H}}^{(n)}$ and ${\mathbb{H}}_{2}={ \mathbb{H}}^{(m-n)}$, ${\rho}_{1}^{-1}\bullet{\rho}$ will be denoted by ${\rho}_{m | n}$.

Now assume that $\sigma$ is a density matrix on ${\mathbb{H}}_{1}$. Then,
$${\rho}(. | {\sigma})=Tr_{1}({{\sigma}\bullet{\rho}_{2 | 1}})$$
is clearly a positive operator on ${\mathbb H}_{2}$. Moreover,
$$Tr({\rho}(. | {\sigma}))=Tr(Tr_{1}({{\sigma}\bullet{\rho}_{2 | 1}}))=Tr(Tr_{1}({\sigma}\bullet{\rho}_{1}^{-1}\bullet{\rho}))=Tr( {\sigma}\bullet{{{\rho}_{1}}^{-1}}\bullet{\rho})$$
$$=Tr( Tr_{2}{({\sigma}\bullet{\rho}_{1}^{-1}\bullet{\rho})})=Tr({\sigma}^{1/2}{\rho}_{1}^{-1/2}(Tr_{2}({\rho})) {\rho}_{1}^{-1/2}{\sigma}^{1/2})$$
$$=Tr({{\sigma}^{1/2}}{{\rho}_{1}^{-1/2}}{{\rho}_{1}}{{\rho}_{1}^{-1/2}}{{\sigma}^{1/2}})={Tr({\sigma})}={1}.$$

Therefore, $ {\rho}(. | {\sigma})$ is a density matrix on ${\mathbb H}_{2}$.

{{\color{black}{Let $\rho \in D(\mathbb{H}^{(n)}\otimes\mathbb{H})$ and $q^{(n)}\otimes q \in Q^{(n+1)}.$ Then $\rho(q\vert q^{(n)})=q(Tr_{1}(q^{(n)}\bullet\rho_{n+1\vert n}))q$ is called the conditional semi-density matrix of q conditioned on $q^{(n)}$ under $\rho$.}}

In the above we assumed that ${\rho}_{1}$ is invertible. For the general case see Lemma \ref{lemNEW4.6_6}.

\begin{definition}
Let $\mathbb{H}$ be a separable Hilbert space and let $ \hat{\rho}=({\hat{\rho}}^{(n)})_{n\in {\mathbb{N}}}$, be a positive operator on ${\mathbb{H}}^{\star}$ and $\bar{\rho}=({\bar{\rho}}^{(n)})_{n\in {\mathbb{N}}}$ where ${\bar{\rho}}^{(n)}= {\hat{\rho}}^{(1)}\bullet{\hat{\rho}}^{(2)}\bullet\cdots\bullet {\hat{\rho}}^{(n)}${\color{black}{,}} be
also a positive operator on ${\mathbb{H}}^{\star}$. Then, the sequence $ \hat{\rho}=({\hat{\rho}}^{(n)})_{n\in {\mathbb{N}}}$ is called a quantum strategy if the sequence
$\bar{\rho}=({\bar{\rho}}^{(n)})_{n\in {\mathbb{N}}}$ is a regular quantum source on ${\mathbb{H}}^{\star}.$
Clearly $ {\hat{\rho}}^{(n+1)} =({\bar{\rho}}^{(n)})^{-1}{\bullet {\bar{\rho}}^{(n+1)}}$ and
$ {\bar{\rho}}^{(n+1)}= {\bar{\rho}}^{(n)}\bullet{\hat{\rho}}^{(n+1)}$
\end{definition}

\begin{lemma}
\label{lem5.8} Let $({\hat{\rho}}^{(n)})_{n\in{\mathbb{N}}}$ be a quantum strategy and $({\bar\rho}^{(n)})_{n\in{\mathbb{N}}}$
be its associated quantum source. Then for each $T\in B(\mathbb{H})$ and each $n\in\mathbb{N}$, $T^{(n)}{\hat{\rho}^{(n)}}={\hat{\rho}^{(n)}}T^{(n)}$ if and only if $T^{(n)}{\bar{\rho}^{(n)}}={\bar{\rho}^{(n)}}T^{(n)}$.

The proof is straightforward.

$\hfill\blacksquare$
\end{lemma}
\begin{remark}
\label{Rem5.3} For future applications we mention that because of the equality ${\hat{\rho}}^{(n+1)}={\bar{\rho}}_{n+1\vert n}$, quantum strategies are also called quantum estimators. Let $({\bar \rho}^{(n)})_{n\in {\mathbb N}}$ be a regular quantum source. It follows from Lemma~\ref{lemNEW4.6_2} and Lemma~\ref{lemNEW4.6_1} that $({{\bar \rho}^{(n)}}_{Q})_{n\in {\mathbb N}}$ is a regular $Q$-quantum source and gives rise to a $Q$-quantum strategy.
\end{remark}

\begin{definition}
\label{def5.7} A quantum estimator $({\hat{\rho}}^{(n)})_{n\in {\mathbb N}}$ is called (weakly, Q-) good with respect to a quantum model $\mathcal{M}$ if its associated quantum source is (weakly, Q-) universal {\color{black}{ relative to}} $\mathcal M$.

Under conditions and notations of Theorems 2 the quantum strategy associated with the weakly universal quantum source ${\bar {\rho}}^{(n)}={\int}_{{\cal M}}{{\beta}(\rho){\rho}^{(n)}}dvol_{\cal M}{(\rho)}$, is weakly good. The quantum strategy associated with the quantum model in Example 3 is also good.

\end{definition}

\begin{example}

Let $\cal M$ be the following quantum model.
$${\cal M}=\{{\rho}_{\theta} |0\leq{\theta}\leq{1} \},$$
where ${\rho}_{\theta}$ is a $2\times 2$-density matrix defined as follows

\[
{\rho}_{\theta} =
\begin{pmatrix}\theta & \sqrt{c(\theta-\theta^2)}\\ \sqrt{c(\theta-\theta^2)} & 1-\theta\end{pmatrix}
\]
and $ 0\leq{c}\leq{1} $ is a real constant.

Let $Q=\{ q_{1} , q_{2} \}$ where $q_{1}=\vert 0\rangle\langle 0\vert$ and $q_{2}=\vert 1\rangle\langle 1\vert$ and $\{\vert 0\rangle , \vert 1\rangle \}$ is the standard basis of the 2-dimensional Hilbert space ${\mathbb{H}}={\mathbb{C}}^{2}.$ Then
\[
{\rho}_{\theta{Q}} = q_{1}{{\rho}_{\theta}}q_{1}+ q_{2}{{\rho}_{\theta}}q_{2}=
\begin{pmatrix}\theta & 0\\ 0 & 1-\theta\end{pmatrix}
\]
is a diagonal matrix.

For simplicity we omit the index $Q$. Assume that $q^{(n)}\in Q^{(n)}$ consists of $k-$times $q_{1}$ and $(n-k)-$times $q_{2}.$ Then for each $0\leq{\theta}\leq{1}$ we have $${q^{(n)}}{{\rho}_{\theta}}^{(n)}{q^{(n)}={\theta}^{k}{(1-{\theta})}^{(n-k)}{q^{(n)}}}.$$ It is straightforward to see that the maximum likelihood estimator for $q^{(n)}$ is
${\rho}_{{\hat{\theta}}(q^{(n)})}$ where ${\hat{\theta}}(q^{(n)})=k/n.$

Clearly ${\cal{M}}$ is a Bayesian quantum model\index{Bayesian quantum model} and its associated quantum source is ${\bar{\rho}}=({\bar{\rho}}^{(n)})_{n\in {\mathbb N}},$ where $ {\bar{\rho}}^{(n)}=
{\int}_{0}^{1}{{{\rho}_{\theta}}^{(n)}}d{\theta}$. Now for $q^{(n)}\in Q^{(n)}$ as above we have
$$ {q^{(n)}}{ \bar{\rho}}^{(n)}{(q^{(n)})}= {\int}_{0}^{1}{ {q^{(n)}} {{\rho}_{\theta}}^{(n)}}{q^{(n)}}d{\theta}
=( {\int}_{0}^{1}{{\theta}^{k}{(1-{\theta})}^{(n-k)}}d{\theta}){q^{(n)}}.$$

One can compute the above integral by partial integration and see that

$$ {q^{(n)}}{ \bar{\rho}}^{(n)}{(q^{(n)})}=\frac{1}{{(n+1)}{n\choose k}}{q^{(n)}}.$$

In the same way for $q^{(n+1)}=q^{(n)}\otimes{q_{1}}\in Q^{(n+1)}$ we have

$${q^{(n+1)}}{ \bar{\rho}}^{(n+1)}{(q^{(n+1)})}=\frac{1}{ {(n+2)} {n+1\choose k+1}}{q^{(n+1)}}.$$
Therefore
$${\hat{\rho}}^{(n+1)}(q_{1}|q^{(n)})=\frac{ {q^{(n+1)}}{ \bar{\rho}}^{(n+1)}{(q^{(n+1)})}}
{ {q^{(n)}}{ \bar{\rho}}^{(n)}{(q^{(n)})}}= \frac{{(n+1)} {n\choose k}}{ {(n+2)} {n+1\choose k+1}}q_{1}=\frac{k+1}{n+2}q_{1}.$$

The density matrix ${\hat{\rho}}^{(n+1)}(.|q^{(n)})$ is called modified maximum likelihood estimator for $q^{(n)}$.
Evidently, for large $n\in{\mathbb N}$ it is very close to ${\rho}_{{\hat{\theta}}(q^{(n)})}$.
\end{example}
Notice that the $Q-$quantum strategy $\hat{\rho}=({ \hat{\rho}}^{(n)})_{n\in{\mathbb N}}$  is not  good.

Let for each $m\in{{\mathbb{N}}}-\{1 \}$,

$${{\cal{M}}}_{m}=\{{{\rho}_{\theta}} | 1/m\leq{\theta}\leq{1-{1/m}} \}.$$

By Theorem 2 the generalized quantum source associated with ${\cal{M}}_{m}$, i.e. ${\bar{\rho}}_{m}= ({\bar{\rho}}_{m}^{(n)})_{n\in {\mathbb N}},$ where $ {\bar{\rho}}_{m}^{(n)}=
{\int}_{1/m}^{1-{1/m}}{{{\rho}_{\theta}}^{(n)}}d{\theta}$, is weakly
universal. Therefore, its associated quantum strategy, i.e. ${\hat{\rho}}_{m}=({\hat{\rho}}_{m}^{(n)})_{n\in{\mathbb{N}}}$ where,
${\hat{\rho}}_{m}^{(n)}=\frac{{\bar{\rho}}_{m}^{(n)}}{{\bar{\rho}}_{m}^{(n-1)}}$
is good. It is straightforward to see that

$$lim_{m\rightarrow{\infty}}{{\bar{\rho}}_{m}}={\bar{\rho}}.$$

and

$$lim_{m\rightarrow{\infty}}{{\hat{\rho}}_{m}}={\hat{\rho}}.$$

\section{Quantum Prediction and Quantum Estimation}

As we said in the introduction, quantum prediction and quantum estimation are the most important subjects of quantum statistical inference. Following the classical works in MDL principle, our method of statistical inference is in general based on universal quantum source and use of it to do quantum prediction and quantum estimation.

\subsection*{Quantum Version of Classical MDL Prediction and Estimation}

Let $\mathbb H$ be a separable Hilbert space and let $Q\in {{\pi}_{0}{(\mathbb H)}}$. Assume that $\cal M$ is a $Q$-quantum model consisting of regular density matrices and for $n\geq{2}$,  ${\hat{\rho}}^{(n)}\in B_{Q+}{(\mathbb H^{(n)})}$ is such that for $I\in {\mathbb{N}}^{(n-1)}$ we have
$$ Tr_1(q_{I}^{(n-1)}\bullet{{\hat{\rho}}^{(n)}})={argmax}_{{\rho}\in{\cal M}}{( q_{I}^{(n-1)}{{\rho}^{(n-1)}} q_{I}^{(n-1)})}, \quad n\geq{2}.$$

Let ${\hat{\rho}}^{(1)}={\rho}_{0}$ be an element of $\cal{M}$. By Lemma~\ref{lemNEW4.6_6}, $\hat{\rho}={({\hat{\rho}}^{(n)})_{n\in {\mathbb N}}}$ is the maximum likelihood $Q$-quantum strategy associated with $\cal M$. In general, $\hat{\rho}$ is not good. But in many cases (see the above example), a modified version of the maximum likelihood $Q$-quantum strategy, which is very close to the unmodified one and the difference between them tends rapidly to zero, is a good one.

This good $Q$-quantum strategy\index{good $Q$-quantum strategy} enables us to predict next outcome given the data $q_{I}^{(n-1)}.$
Moreover, let the data $q_{I}^{(n-1)}$ be really generated by ${\rho}\in \mathcal{M}.$ Then as we will see in the next chapter $ Tr_1(q_{I}^{(n-1)}\bullet{{\hat{\rho}}^{(n)}})$ can be considered as an estimation of $\rho$.

\subsection*{Quantum Version of Classical two-part code estimation}

Let $\mathbb H$ be a separable Hilbert space and let $Q\in {{\pi}_{0}{(\mathbb H)}}$. Assume that $\underline{\mathcal M}$ is a generalized quantum model. For $I\in {\mathbb{N}}^{(n)}$, let $\ddot{\rho}_n$ be defined as follows
$$\ddot{\rho}_n=\omega({argmax}_{{{\rho}}\in{\underline{\mathcal{M}}}}{q_{I}^{n}{{\rho}}^{(n)}{q_{I}^{n}}}).$$

If the maximum is achieved by more than one $\rho$ we choose the one with the maximum trace. And if there is still more than one $\rho$ there is no further preference.
More precisely, let us suppose that $\underline{\mathcal{M} }$ is a compact Riemannian sub-manifold of the Hilbert space $(B_{T}(\mathbb H) , {\langle . \vert . \rangle}_{T} )$ consisting of semi-density matrices, where for $\rho$ and ${\rho}^{\prime}$ in $B_{T}(\mathbb H)$, $\langle {\rho} \vert {\rho}^{\prime}\rangle ={\rm Tr}({\rho}{\rho}^{\prime})$ and  $(\underline{\mathcal{M} }, {\Sigma}, {\mu})$ is its associated canonical measure space. To obtain $\ddot{\rho}_n$, let $Z$ be the set of all extremum points of the smooth function $h:{\rho}\longrightarrow {{\rm Tr}(q_{I}^{n}{\rho}^{(n)}{q_{I}^{n}})}$ on $\underline{\mathcal{M} }$, and let $Z^{\prime}$ be the set of all elements ${\rho}\in Z$ at which the bundle map
$Hessian(h): T{\underline{\mathcal{M} }}\longrightarrow{T{\underline{\mathcal{M} }}}$ is negative definite. Clearly, all points of $Z^{\prime}$ are maximum points of $h$. Therefore,

$$\ddot{\rho}={\omega}(argmax_{{\rho}\in{Z^{\prime}}}{h(\rho)}).$$

In the next section we will show that given the outcome $q_{I}^{(n)}$, $\ddot{\rho}_n$ is an estimator of the state of the system.

\begin{example}
\label{ex8.1}\hfill

\noindent
Let the quantum exponential family $(\rho_t)_{t\in \mathbb{R}}$ be defined as follows $$ \rho_t=e^{\frac{1}{2}[t\sigma_z-\gamma(t)]}\rho_0e^{\frac{1}{2}[t\sigma_z-\gamma(t)]},$$ where $${\sigma}_{z}=
\dfrac{1}{2}\begin{bmatrix}
1 & 0 \\
0 & -1 \\
\end{bmatrix},\rho_0=
\dfrac{1}{2}\begin{bmatrix}
1 & x_0 \\
x_0 & 1 \\
\end{bmatrix}, -1\leq x_0 \leq 1, \gamma(t)=\log[\text{Tr}( \rho_0e^{t\sigma_z})].$$

now we perform measurement on the state space $\mathbb{H}$, the two dimensional Hilbert space, by the system of measurement $ \Pi = \lbrace q_{0}=\vert 0\rangle\langle 0 \vert,q_{1}=\vert 1\rangle\langle 1 \vert\rbrace$ n times and obtain ${n_0}$ times $q_{0}$ and ${n_1}$ times $q_{1}$. We want to estimate the state of the system.
Let ${\hat{\gamma}}(\rho)={\nu}(\rho)+{\gamma}(\rho)$ where ${\nu}(\rho)$ is the $Q$-quantum complexity of $\rho$, ${\gamma}(\rho)$ is its $Q$-Shannon entropy and let $\rho_T , T=argmax_{t}{\{e^{-{\hat{\gamma}}(\rho)}(q_0\rho_t q_0)^{n_0}(q_1\rho_t q_1)^{n_1}}\}$ be the estimator obtains by our method. Clearly

\begin{equation}
\begin{array}{rl}
T&=argmax_{t}{\{e^{-{\hat{\gamma}}(\rho_t)}(q_0\rho_t q_0)^{n_0}(q_1\rho_t q_1)^{n_1}}\} \\
&=argmax_{t}\lbrace{-({\nu(\rho_t)}+{\gamma(\rho_t)})+t(n_0-n_1)}-n\ln({e^t+e^{-t}})\rbrace\\
\end{array}
\end{equation}
where,
\begin{equation}
\begin{array}{rl}
\rho_t&=
\dfrac{1}{e^t+e^{-t}}\begin{bmatrix}
e^t & x_0 \\
x_0 & e^{-t} \\
\end{bmatrix} \\
\gamma(\rho_t) &=\dfrac{e^t}{e^t+e^{-t}}\ln({\dfrac{e^t}{e^t+e^{-t}}})+\dfrac{e^{-t}}{e^t+e^{-t}}\ln({\dfrac{e^{-t}}{e^t+e^{-t}}})\\
&=t \tanh(t) -\ln(e^t+e^{-t})\\
\nu(\rho_t) &=\dfrac{x_0}{e^t+e^{-t}}
\end{array}
\end{equation}
Therefore,

$$T=argmax_{t}{-({\dfrac{x_0}{e^t+e^{-t}}}+{t \tanh(t) -\ln(e^t+e^{-t})})+t(n_0-n_1}-n\ln({e^t+e^{-t}}).$$
equivalently,
$$\dfrac{d\bigg({-({\dfrac{x_0}{e^t+e^{-t}}}+{t{\tanh(t)}-\ln(e^t+e^{-t})})+t(n_0-n_1)}-n\ln({e^t+e^{-t}})\bigg)}{dt}\bigg\vert_{t=T}=0$$
or $$(k-n)y^4+x_0 y^3+2ky^2-x_0 y-4y^2\ln y+k+n=0$$ where $n_0-n_1=k$ and $e^T=y$.

It is easy to see that the best estimation according to the MLE is $y=\sqrt{\dfrac{n+k}{n-k}}$, which doesn't depend on $x_0$ and doesn't get any information about it.
As the following table shows the estimator obtained by our method is eventually the same as the ML estimator.
\begin{center}
\begin{tabular}{||c c c c c c c||}
\hline
n & $n_0$ & $n_1$ & k & $x_0$ & MLE results & Our methods \\ [0.5ex]
\hline\hline
10 & 8 & 2 & 6 & 0.75 & y=2 & y=1.92165 \\
\hline
& & & & 1 & y=2 & y=1.93858 \\
\hline
& & & & 0 & y=2 & y=1.87383 \\
\hline
100 & 80 & 20 & 60 & 0.75 & y=2 & y=1.99180 \\
\hline
& & & & 1 & y=2 & y=1.99366 \\
\hline
& & & & 0 & y=2 & y=1.98627 \\
\hline
100 & 5 & 95 & -90 & 0.5 & y=0.2316 & y=0.2294 \\
\hline
1000 & 560 & 440 & 120 & 0.5 & y=1.1281 & y=1.1280 \\
\hline
& & & & 1 & y=1.1281 & y=1.1280 \\
\hline
& & & & 0 & y=1.1281 & y=1.1280 \\ [1ex]
\hline
\end{tabular}
\end{center}

\end{example}

\section{Consistency and Convergence}

Consistency is a very important property of different methods of statistical (inductive) inference. Let us explain briefly what we mean by it.

Assume that $\mathbb H$ is a separable Hilbert space and $\mathcal M$ is a quantum model on $\mathbb H$. we say that a method of quantum statistical inference is consistent with respect to $\mathcal{M}$ if for $\rho_{0}\in \mathcal{M}$ and $Q\in\pi_0({\mathbb{H}})$, we perform the quantum measurement $Q$ on the quantum system $\mathbb{H}$ in the state $\rho_0$ repeatedly and obtain more and more data the state yielded by the method is more and more close to the state $\rho_0$ in some sense.

we emphasize that the above definition of consistency depends on the quantum model $\cal{M}$ and on
$Q\in\pi_0({\mathbb{H}})$.

In this section we investigate different approaches to consistency and convergence.

\subsection{Consistency based on distinguishability\index{distinguishability} }

\begin{convention}
Let $X$ be a complex vector space and let ${\lambda}\in{\mathbb{C}}$. Then

1) According to the situation the same letter $\lambda$ also denotes the constant function
$X:\longrightarrow{\{{\lambda}\}}{\subset}{\mathbb{C}}.$

2)Let $T$ and $S$ be in $X$. Assume that $T={\lambda}S$. Then, we put  $T/S={\lambda}.$

\end{convention}

Assume that $\mathbb H$ is a separable Hilbert space and $Q\in{{\pi}_{0}{(\mathbb H)}}$. Let ${\bar{\rho}}=({\bar{\rho}}^{(n)})_{n\in{\mathbb N}}$ be a quantum source on ${\mathbb H}^{\ast}.$ For each $n\in{\mathbb N}$ let $P_{n}$ be a unary relation on $Q^{(n)}.$ Then,
$$Tr({\sum}_{q^{(n)}\in{Q^{(n)}}\vert P_{n}{(q^{(n)})}}{q^{(n)}{\bar{\rho}}^{(n)}q^{(n)}})$$
\noindent
will be denoted by ${\bar{\rho}}(P_{n})$. suppose that ${\bar{\rho}}^{\prime}=({{\bar{\rho}}^{\prime{(n)}}})_{n\in {\mathbb N}}$ is another quantum source on ${\mathbb H}^{\ast}$. For each $n\in {\mathbb N}$, and each ${\delta}> {0}$ let
${P_{n}}^{\delta}$ be the unary relation

$$\frac{q^{(n)}{{\bar{\rho}}^{\prime{(n)}}}q^{(n)}}{q^{(n)}{{\bar{\rho}}^{(n)}}q^{(n)}}>{\delta}$$

on $Q^{(n)}$.
\begin{definition}
\label{def8.1.1} Under the above notations and conventions we say that, ${\bar{\rho}}^{\prime}$ is asymptotically distinguishable from $\bar{\rho}$ if for all ${\delta}> 0$ we have

$$lim_{n\rightarrow {\infty}}{{\bar{\rho}}{(P_{n}^{\delta})}}=0.$$
\end{definition}

{\color{black}{Let $\underline{\mathcal M}$ be a countable Bayesian set of regular generalized quantum sources on ${\mathbb H}^{\star}$ and $\cal M$ be its associated set of quantum sources. For each $n\in{\mathbb N}$, let us denote $\omega({\bar{\rho}}^{(n)})$ by $ {\rho}^{(n)} $. For each $q^{(n)}\in Q^{(n)}$ define ${\ddot{\rho}}_{(n)}$ as follows:     
   $$ (\star)  {{\ddot{\rho}}_{(n)}}= argmax_{{\rho}\in{{\cal{M}}}}{q^{(n)}{\bar{\rho}}^{(n)}q^{(n)}}.$$

Observe that $\ddot{\rho}_{(n)}$ depends on $q^{(n)}$. }}

Now we have the following important consistency theorem.

\begin{theorem}
\label{theo4.1} Let $\mathbb{H}$, $Q$, $\underline{\cal{M}}$ , $\mathcal {M}$ and $\ddot{\rho}_{(n)}$ be as above. Let $ {\color{black}{{\bar{\rho}}^{\ast}}} \in{{\underline{\cal{M}}}}$ and $\ddot{\cal{M}}$ be the subset of ${\cal{M}}$ consisting of quantum sources asymptotically distinguishable from ${\rho}^{\ast}$. Then
$$lim_{n\rightarrow {\infty}}{{\rho}^{\ast}{({\ddot{\rho}_{(n)}}\in{{\ddot{\cal{M}}}})}}=0.$$
\end{theorem}

\noindent{\bf{Proof}.}\quad The proof is the same as the proof of Theorem 5.1 of [4] with necessary modifications.

{\color{black}{Let ${\ddot {\rho}}_{(n)}\in {\ddot{\cal{M}}}$. From the equality $(\star)$ , for some $ {{\bar{\rho}}}\in \underline{\cal{M}}$ we have
$$q^{(n)}{{\bar{\rho}}^{(n)}}q^{(n)}\geq q^{(n)}{{\bar{\rho}}^{\ast{ (n) }}}q^{(n)}$$

Therefore, for each subset ${\cal{M}}^{\prime}$ 
of $\ddot{\cal{M}}$ we have,
$${{\rho}^{\ast}{({\ddot{\rho}_{(n)}}\in{{\cal{M}}^{\prime}})}}\leq{\rho}^{\ast}{\{for some {\tau}\in {\cal{M}}^{\prime},\frac{q^{(n)}{\tau}^{(n)}q^{(n)}}{q^{(n)}{{\rho}^{\ast(n)}}q^{(n)}}\geq
\frac{Tr({\bar{\rho}}^{\ast})}{Tr{(\bar{\tau})}} \}}.$$

Let us denote $\frac{Tr( {\bar{\rho}}^{\ast} )}{Tr(\bar{\tau})}$ by ${\delta}(\tau)$. Assume that $n:\rightarrow{{{\rho}}_{n}}$ is a bijective mapping from $\mathbb N$ onto
${\ddot{{\cal {M}}}}$ and $m=Tr(\sum_{n=1}^{\infty}{{\bar{\rho}}_{n}}).$
}}
Let ${\epsilon}>{0}$ be given and let ${\pi}=m-{\epsilon}{Tr({\bar{\rho}}^{\ast})}.$ Suppose that $N$ is the least integer such that $Tr(\sum_{n=1}^{N}{{\bar{\rho}}_{n}})\geq{\pi}.$ Let
${\ddot{\bar{\cal M}}}=\{{\rho}_{n} | 1\leq{n}\leq{N} \}$ and
$\bar{\ddot{\cal M}}= {\ddot{\cal M}}-{{\ddot{\bar{\cal M}}}}$.

Evidently,$$ {{\rho}^{\ast}{({\ddot{\rho}_{(n)}}\in{\ddot{\cal M}})}}
\leq
{{\rho}^{\ast}{({\ddot{\rho}_{(n)}}\in{{\ddot{\bar{\cal M}}}})}}+
{{\rho}^{\ast}{({\ddot{\rho}_{(n)}}\in{\bar{\ddot{\cal M}}})}}.$$
and
$$\lim_{n\rightarrow{\infty}}{{{\rho}^{\ast}{({\ddot{\rho}_{(n)}}\in{\ddot{\cal M}})}}}
\leq \lim_{n\rightarrow{\infty}}{{\rho}^{\ast}{({\ddot{\rho}_{(n)}}\in{{\ddot{\bar{\cal M}}}})}}+\lim_{n\rightarrow{\infty}}{{\rho}^{\ast}{({\ddot{\rho}}_{(n)}\in{\bar{\ddot{\cal M}}})}}.$$

Assume that ${\rho}\in{\ddot{\cal M}}$ and ${\delta}(\rho)= Tr({\bar{\rho}}^{\ast})/Tr(\bar{\rho})$. Since ${\rho}$ is asymptotically distinguishable from
${\rho}^{\ast}$, $\lim_{n\rightarrow{\infty}}{{\rho}^{\ast}{(P_{n}^{\delta(\rho)})}}=0$. Since ${{\ddot{\bar{\cal M}}}}$ is a finite set we have

$$\lim_{n\rightarrow{\infty}}{{\rho}^{\ast}{({\ddot{\rho}_{(n)}}\in{\ddot{\bar{\cal{M}}}})}}\leq{\lim_{n\rightarrow{\infty}}{\sum_{{\rho}\in{{\ddot{\bar{\cal {M}}}}}}{{\rho}^{\ast}{(P_{n}^{\delta{(\rho)}})}}=\sum_{{\rho}\in{\ddot{\bar{M}}}}{ \lim_{n\rightarrow{\infty}}{{\rho}^{\ast}{(P_{n}^{\delta{(\rho)}}})}}}}=0.$$

On the other hand by the fundamental coding theorem we have

$${\rho}^{\ast}{(P_{n}^{\delta(\rho)})}\leq{1/{\delta(\rho)}}. $$
Hence,
\begin{equation*}
\begin{array}{rl}
{{\rho}^{\ast}{({\ddot{\rho}_{(n)}}\in{\bar{\ddot{{\cal M}}}})}}&\leq
{\sum_{{\rho}\in{\bar{\ddot{\cal M}}}}{{\rho}^{\ast}}{(P_{n}^{\delta{(\rho)}})}}\\
&\leq{{\sum_{{\rho}\in{\bar{\ddot{\cal M}}}}{1/{\delta(\rho)}}}}\\
&=\sum_{{\rho}\in{\bar{\ddot{\cal M}}}}{{Tr(\bar\rho)}/{Tr({\bar\rho}^{\ast})}}\\
&\leq(m-{\pi})/{Tr({\bar\rho}^{\ast})}\\
&={\epsilon}.
\end{array}
\end{equation*}
Therefore,
$$\lim_{n\rightarrow{\infty}}{{\rho}^{\ast}{({\ddot{\rho}_{(n)}}\in{\bar{\ddot{\cal M}}})}}=0.$$

$\hfill\blacksquare$
\bigskip

{\color{black}{
\subsection{Consistency in terms of KL risk and Cezaro average KL risk}
\index{KL risk} \index{Cezaro average KL risk}
\begin{theorem}
\label{theo8.2.1} Let $\bar {\rho}$ and ${\rho}^{\ast}$ be regular quantum sources and ${\rho}^{\ast}$ be simple. Then
$$S({{\rho}^{\ast}}^{(n)}\Vert{\bar{\rho}}^{(n)})=^{w}\sum_{i=1}^{n}{E_{{{\rho}^{\ast}}^{(i-1)}} S({\rho}^{\ast}_{i\vert(i-1)}\Vert{{\bar{\rho}}_{i\vert(i-1)}})}.$$
\end{theorem}

\noindent{\bf{Proof}.}\quad Assume that $Q$ is a complete set of mutually orthogonal minimal projections. By Lemma~\ref{lemNEW4.6_2} ${\rho}^{\ast(n)}_{Q^{(n)}}$ and ${\bar {\rho}}^{(n)}_{Q^{(n)}}$ are invertible. For simplicity we omit the subscript $Q^{(k)}$.
By definition and previous lemmas and theorems we have:

\begin{equation}
\begin{array}{rl}
S({{\rho}^{\ast}}^{(n)}\Vert{\bar{\rho}}^{(n)})&={\rm Tr}({{\rho}^{\ast}}^{(n)}\log{{\rho}^{\ast}}^{(n)}-
{{\rho}^{\ast}}^{(n)}\log{{\bar{\rho}}^{(n)}}) \\
&={\rm Tr}{{{\rho}^{\ast}}^{(n)}}(\log{{\rho}^{\ast}}^{(n)}-\log{{\bar{\rho}}^{(n)}})\\
&={\rm Tr}(({\Pi}_{i=1}^{n}({\rho}^{\ast}_{i\vert(i-1)})) (\log{{\Pi}_{i=1}^{n}}{\rho}^{\ast}_{i\vert(i-1)}- \log{{\Pi}_{i=1}^{n}}{{\bar{\rho}}_{i\vert(i-1)}}))\\
&=\sum_{i=1}^{n}{\rm Tr}({{{\rho}^{\ast}}^{(i-1)}}
{\rho}^{\ast}_{i\vert(i-1)}
(\log{\rho}^{\ast}_{i\vert(i-1)}-\log{{\bar{\rho}}_{i\vert(i-1)}}))\\
&=\sum^{n}_{i=1} E_{{{\rho}^{\ast}}^{(i-1)}} S({\rho}^{\ast}_{i\vert(i-1)}\Vert{{\bar{\rho}}_{i\vert(i-1)}}).
\end{array}
\end{equation}
(See also [4].)

$\hfill\blacksquare$

\begin{definition}
Let ${\rho}^{\ast}$ and ${\bar{\rho}}$ be regular quantum sources and $\hat{{\rho}^{\ast}}$ and $\hat{\rho}$ be their associated quantum strategies. Moreover, let ${\rho}^{\ast}$ be simple. Then, the $n-$th order standard KL-risk of
${\hat{{\rho}^{\ast}}}$ with respect to ${\hat{\rho}}$ is

$$RISK_{n}(\hat{{\rho}^{\ast}}, {\hat{\rho}})=^w E_{{\rho^{\ast}}^{(n-1)}}{[S({\hat{{\rho}^{\ast}}}^{(n)}\Vert{\hat{\rho}}^{(n)})]}.$$
And the $n-$th order Cezaro average risk of ${\hat{{\rho}^{\ast}}}$ with respect to ${\hat{\rho}}$ is
$$\bar{RISK_{n}}{(\hat{{\rho}^{\ast}}, {\hat{\rho}})}=1/n {S({{\rho}^{\ast}}^{(n)}\Vert{\bar{\rho}}^{(n)})}=1/n{\sum_{i=1}^{n}{{RISK}_{i}}}{ (\hat{{\rho}^{\ast}}, {\hat{\rho}})}.$$
\end{definition}
\begin{theorem}(Convergence Theorem for quantum Estimators)
\label{theo8.2.2} Let $\mathbb{H}$ be a separable Hilbert space and $Q\in{{\pi}_{0}{(\mathbb H)}}.$ Let $\cal{M}$ be a quantum model on ${\mathbb{H}}$ and $\bar{\rho}$ be a reqular $Q$-universal quantum source with respect to $\cal{M}$. Then $\hat{\rho}$ the $Q$-quantum estimator associated with $Q$-universal quantum source ${\bar {\rho}}_{Q}$ is Cezaro consistent with respect to
$\mathcal{M}$. In other words for all ${\rho}^{\ast}\in{\cal{M}}$ we have
$$\lim_{n\rightarrow{\infty}}{1/n{\sum_{i=1}^{n}{{RISK}_{i}}}{ (\hat{{\rho}^{\ast}}, {\hat{\rho}})}}=0.$$
The proof is a consequence of the definition of $Q$-universal source and Theorem \ref{theo8.2.1}.
$\hfill\blacksquare$
\end{theorem}

\begin{lemma}
\label{cons1.1}
Let $f$ and $F$ be two increasing positive real functions defined on ${\mathbb R}^{+}.$ If the function $f/F$ is decreasing and $ f=O(F)$, then $f(n+1)-f(n)=O(F(n+1)-F(n)).$
\end{lemma}
{\bf{Proof}.} Assume that there exists $c> {0}$ such that for $n$ large enough $f(n)\leq{cF(n)}.$ Let $f(n+1)=c_{1}{F(n+1)}$ and $f(n)=c_{0}{F(n)}$. Then,

$$ f(n+1)-f(n)= c_{1}{F(n+1)}- c_{0}{F(n)}.$$ Since $ c_{1}\leq{ c_{0}}\leq{c}$ we have
$$ f(n+1)-f(n)= c_{1}{F(n+1)}- c_{0}{F(n)}\leq{ c_{0}({F(n+1)}-{F(n)})}\leq{ c({F(n+1)}-{F(n)})}.$$
Therefore, $f(n+1)-f(n)=O(F(n+1)-F(n)).$

\begin{lemma}
\label{cons1.2}
Let $f:{\mathbb R}^{+}\longrightarrow {\mathbb R^{+}}$ be a differentiable decreasing function, and ${\int}_{0}^{1/2}{f(x)dx}<{\infty}$. Let $F(x)={\int}_{0}^{x}{f(x)dx}$, 
and let
$(a_{n})_{n\in{\mathbb N}}$ be a sequence of non-negative real numbers. 
Then

1) If $a_{n}=O(f(n)),$ Then ${{\sum}_{i=1}^{n}{a_{i}}}=O(F(n))$. Conversely, if for $n$ large enough the function ${{\sum}_{i=1}^{n}{a_{i}}}/(F(n+1))$ is decreasing and ${{\sum}_{i=1}^{n}{a_{i}}}=O(F(n))$, then $a_{n}=O(f(n))$.

2) If $lim_{n\rightarrow {\infty}}{a_{n}}={0}$, then  $lim_{n\rightarrow {\infty}}{{\sum}_{i=1}^{n}{a_{i}}}/n={0}$. Conversely, if for $n$ large enough the function ${{\sum}_{i=1}^{n}{a_{i}}}/n$ is decreasing and $lim_{n\rightarrow{\infty}}{{\sum}_{i=1}^{n}{a_{i}}}/n=0$, then $ lim_{n\rightarrow{\infty}}{ a_{n}}=0$.

\end{lemma}

\noindent{\bf{Proof}.}\quad

1) In approximating the integral by sum and remembering the fact that the function $f$ is decreasing,  for $1\leq n\in{\mathbb{N}}$ we have $0\leq{{\int}_{0}^{n}{f(x)dx}-{{\sum}_{i=1}^{n}{f(i)}}}$, and

$${\int}_{0}^{n}{f(x)dx}-{{\sum}_{i=1}^{n}{f(i)}}\leq{{\int}_{0}^{1/2}{f(x)dx}}+1/2[f(1/2)-f(n)]
\leq{{\int}_{0}^{1/2}{f(x)dx}}+1/2[f(1/2)].$$

Therefore,

$$ F(n)={{\sum}_{i=1}^{n}{f(i)}}+O(1).$$
Hence,

$${\sum}_{i=1}^{n}{a_{i}}=O({\sum}_{i=1}^{n}f(i))
=O(F(n)+O(1))=O(F(n)).$$

Conversely, assume that ${{\sum}_{1}^{n}{a_{i}}}=O(F(n))$. Since $F$ is increasing
${{\sum}_{1}^{n}{a_{i}}}=O(F(n+1))$. Therefore, there exists a constant ${0<c}\in {\mathbb{R}}$ such that for all $n\in {\mathbb{N}}$ greater than some $n_{0}$ we have
${{\sum}_{1}^{n}{a_{i}}}\leq{cF(n+1)}$. By the above lemma we have

$$a_{n}= {\sum}_{1}^{n}{a_{i}}-{\sum}_{1}^{n-1}{a_{i}}\leq{c(F(n+1)-F(n))}=cf({\theta}_{n}).$$

Where, ${n}\leq{{\theta}_{n}}\leq{n+1}.$ Since $f$ is decreasing we have
$f({\theta}_{n})\leq{f(n)}$. Therefore, $ a_{n}=O(f(n)).$

2) From the equality $lim_{n\rightarrow{\infty}}{a_{n}}={0}$ it follows that for each ${\epsilon}> 0$ there exists $n_{1}\in{\mathbb{N}}$ such that for all $n\geq{n_{1}}$, we have $ a_{n}\leq{\epsilon}.$ Suppose that for $k\in{\mathbb{N}}$, ${\frac{\sum_{i\leq{n_{1}}} {a_{i}}}{k}}\leq{\epsilon}$. Let $ n\in{\mathbb{N}}$  be such that ${k-2{n_{1}}}\leq{n}.$ Then
$2(n+n_{1})\geq{k+n}.$ So,
$$\sum_{i=1}^{n+n_{1}}{a_{i}}=\sum_{i=1}^{n_{1}}{a_{i}}+\sum_{i=n_1+1}^{n+n_1}{a_{i}}\leq{(k+n){\epsilon}}
\leq{2(n+n_{1}){\epsilon}}.$$

Hence,$ \frac{\sum_{i=1}^{n_{0}}{a_{i}}}{ n_{0}}\leq{2{\epsilon}}$, where $n_{0}=n+n_{1}.$ It is clear that for all $n\geq{n_{0}}$ we have
$$ \frac{\sum_{i=1}^{n}{a_{i}}}{ n}\leq{2{\epsilon}}.$$
Therefore,
$$lim_{n\rightarrow{\infty}}{ \frac{\sum_{i=1}^{n}{a_{i}}}{ n}}= 0.$$

Conversely, since for $n$ large enough the sequence $\frac{\sum_{i=1}^{n}{a_{i}}}{ n}$ is decreasing we have $ \frac{\sum_{i=1}^{n+1}{a_{i}}}{ n+1}\leq {\frac{\sum_{i=1}^{n}{a_{i}}}{ n}}.$ So,
$$ {\sum_{i=1}^{n+1}{a_{i}}}\leq\dfrac{(n+1)}{n}{ \sum_{i=1}^{n}{a_{i}}}.$$
Therefore, $a_{n+1}\leq {\frac{\sum_{i=1}^{n}{a_{i}}}{ n}}.$ But   $lim_{n\rightarrow{\infty}}{{\frac{\sum_{i=1}^{n}{a_{i}}}{ n}}}={0}$. Therefore,
 $lim_{n\rightarrow{\infty}}{a_{n}}={0}.$ 

\begin{theorem}
Let $\mathbb{H}$ be a separable Hilbert space. Assume that ${\rho}^{\ast}=({{\rho}^{\ast}}^{(n)})_{n\in{\mathbb{N}}}$ and ${\bar{\rho}}=({{\rho}}^{(n)})_{n\in{\mathbb{N}}}$ are quantum sources on the space ${\mathbb{H}}^{\ast}$ and ${\rho}^{\ast}$ is simple. Then

1) if $lim_{n\rightarrow{\infty}}{RISK_{n}(\hat{\rho}^{{\ast}}, \hat{{\rho}})}=^{w}0$ then
$lim_{n\rightarrow{\infty}}{ \frac{1}{n}S({\rho}^{{\ast(n)}}\Vert {\bar{\rho}}^{(n)})}= ^{w}{0}.$ Conversely, if for large $n$ , $ \frac{1}{n}S({\rho}^{{\ast(n)}}\Vert {\bar{\rho}}^{(n)})$ is decreasing and
$lim_{n\rightarrow{\infty}}{ \frac{1}{n}S({\rho}^{{\ast(n)}}\Vert {\bar{\rho}}^{(n)})}= ^{w}{0}$
then $lim_{n\rightarrow{\infty}}{RISK_{n}(\hat{\rho}^{{\ast}}, {\hat{\rho}})}=^{w}{0}$

2) Let $f:{\mathbb R}^{+}\longrightarrow {\mathbb R^{+}}$ be a differentiable decreasing function, and ${\int}_{0}^{1/2}{f(x)dx}<{\infty} $. Let $F(x)= {\int}_{0}^{x}{f(t)dt}.$
Then, if ${RISK_{n}(\hat{\rho}^{{\ast}}, {\hat{\rho}})}=^w O(f(n))$ then
$$ S({\rho}^{{\ast(n)}}\Vert {\bar{\rho}}^{(n)})=O(F(n)).$$

Conversely, if $\frac{ S({{{\rho}^{\ast}}^{(n)}}\Vert {\bar{\rho}}^{(n)})}{F(n+1)} $ is decreasing and
$ S({{{\rho}^{\ast}}^{(n)}} ||{\bar{\rho}}^{(n)})={O(F(n))}$, then
${RISK_{n}(\hat{\rho}^{{\ast}}, {\hat{\rho}})}=^w O(f(n))$
\end{theorem}
\noindent\textbf{Proof.}\quad The proof is a consequence of the definitions and Lemma \ref{cons1.2}. See also [4].

$\hfill\blacksquare$
}}

\subsection{Consistency in terms of Renyi divergences and Hellinger distance}
\index{Renyi divergences} \index{Hellinger distance}
Let $\mathbb H$ be a Hilbert space. Let ${\rho}_{1}$ and ${\rho}_{2}$ be density matrices. Then

1) The natural quantum relative entropy of ${\rho}_{1}$ to ${\rho}_{2}$ is
$$S_{nat}({\rho}_{1}\Vert {\rho}_{2})=Tr({\rho}_{1}{ln{{\rho}_{1}}})-Tr({\rho}_{1}{ln{{\rho}_{2}}}).$$

2) The Helinger distance of ${\rho}_{1}$ and ${\rho}_{2}$ is
$$He^{2}( {\rho}_{1} || {\rho}_{2})=|| {\rho}_{1}^{1/2}-{\rho}_{2}^{1/2}||_{T}^{2}$$

3) Let $1>{\lambda}>0$ be a real number. The Renyi divergence of order $\lambda$ of ${\rho}_{1}$ and ${\rho}_{2}$ is defined as follows:
$${\bar{d}}_{\lambda}{({\rho}_{1} ||{\rho}_{2})}=-{\frac{1}{1-{\lambda}}{ln(<{\rho}_{1}^{\lambda} |
{{\rho}_{2}}^{1-{\lambda}}>_{T})}}.$$

Observe that

$$He^{2}( {\rho}_{1} || {\rho}_{2})=|| {\rho}_{1}^{1/2}-{\rho}_{2}^{1/2}||_{T}^{2}=
Tr[( {{\rho}_{1}^{1/2}-{\rho}_{2}^{1/2}})^{2}]=Tr({\rho}_{1}+{\rho}_{2}-2{{\rho}_{1}}^{1/2}{{\rho}_{2}}^{1/2})$$
$$=2(1-Tr({{\rho}_{1}}^{1/2}{{\rho}_{2}}^{1/2}) ) \leq{[-2ln<{\rho}_{1}^{1/2}|{\rho}_{2}^{1/2}>_{T}]}
={{\bar{d}}_{1/2}( \rho}_{1} || {\rho}_{2}). $$

{\color{black}{

Assume that the Hilbert space $\mathbb{H}$ is the state space of a quantum system. let ${\cal{M}}=\{{\tau}_{n} |  n\in{\mathbb{N}} \}$ be a countable quantum model and let $(u_{n})_{ n\in {\mathbb{N}}}$ be a sequence of nonzero positive real numbers such that
${\sum}_{ n\in {\mathbb{N}}}{u_{n}}=1$. The set consisting of all elements of the form ${ u_{n}}{{\tau}_{n}}$ will be denoted by ${\underline{\cal{M}}}$. Let $Q\in{\pi}_{0}{(\mathbb{H})}.$
For $\alpha \geq 1$, let ${\underline{\cal M}}_{{\alpha}}=\{ {\rho}_{\alpha} | {\rho}\in {\underline{\cal M}} \}$. Where, ${\rho}_{\alpha}=[Tr(\rho)]^{{\alpha}-1}{\rho}.$ Let  ${\bar{{\rho}_{\alpha}}}^{(n)}$ be defined as follows:

$$  {q^{n}}{\bar{{\rho}_{\alpha}}}^{(n)}{q^{n}}=max_{{\rho}_{\alpha}\in{\underline{\cal M}_{\alpha}}}( {q^{n}{{\rho}_{\alpha}}^{(n)}{q^{n}}}).$$
Assume that $({\bar{{\rho}_{\alpha}}}^{(n)})_{n\in {\mathbb N}}$ is a universal semi-density matrix for ${\cal M}.$ Suppose  ${{{\ddot{\rho}}_{ n}}}$ is defined as follows:

For $q^{(n)}\in Q^{(n)}$ ,
$${{\ddot{\rho}}_{n}}={\omega}(argmax_{{\rho}\in{\underline{\cal M}}}{q^{(n)}{{ \rho}^{(n)}} q^{(n)})}.$$
Observe that $\ddot{\rho}_{n}$ depends on $q^{(n)}$.
Let ${\rho}_{k}={u_{k}}{\tau}_{k}=argmax _{{\rho}\in{\underline{\cal M}}} {q^{(n)}{{ \rho}^{(n)}} q^{(n)}}.$ Then evidently

$$ {q^{(n)}} {\bar{{\rho}_{\alpha}}}^{(n)}{q^{(n)}}= {u_{k}}^{\alpha} {q^{(n)}} {\ddot{\rho}}_{n}^{(n)}{q^{(n)}}.$$
In the following, we write ${\ddot{\rho}}_{n}$ instead of ${\ddot{\rho}}_{n}^{(n)}.$}}

\begin{theorem}
\label{theo8.2.3} Let ${\rho}^{\ast}$ be the state of the system. Under the above notations and conventions for all $\alpha>1$ and ${0}< {\lambda}={1-1/{\alpha}}$ we have

$$E_{\rho^{\ast(n)}_{Q}}{({\bar{d}}_{\lambda}{(\rho^{\ast(n)}_{Q}\Vert {\ddot{\rho}}_{nQ}^{(n)})})}\leq{{\frac{1}{n}}S_{nat}(\rho^{\ast(n)}_{Q}\Vert {{\bar{\rho}}_{{\alpha} Q}^{(n)}}}).$$

And for $\alpha=2$ we have

$$E_{\rho^{\ast(n)}_{Q}}{({{He}^{2}{(\rho^{\ast(n)}_{Q}\Vert \ddot{\rho}_{ nQ}^{(n)})}})}\leq{{\frac{1}{n}}S_{nat}(\rho^{\ast(n)}_{Q}\Vert {{\bar{\rho}}_{\alpha Q}^{(n)}}}).$$
\end{theorem}

\noindent{\bf{Proof}.} (The proof is a modified version of the proof of Theorem 15.3 of [4].)

For simplicity we omit the index $Q$. Since  $\lambda = 1-{\alpha}^{-1}.$ we have $\alpha = 1/{1-\lambda}.$ Let $A({\rho}^{\ast} || {\ddot{\rho}}) = Tr({{\rho}^{\ast}}^{\lambda}{ \ddot{\rho}}^{1-{\lambda}}).$ For each $q^{(n)}\in Q^{(n)}$ we have

\begin{equation*}
\begin{array}{rl}
{\color{black}{q^{(n)}}}{\bar{d}}_{\lambda}{({{\rho}^{\ast}}^{(n)}\Vert
{\color{black}{{\ddot{\rho}}_{n}}})}&=(-1/{1-\lambda}){\color{black}{q^{(n)}}}{{{\color{black}{\ln}}}
A({{\rho}^{\ast} }^{(n)}||
{\color{black}{{\ddot{\rho}}_{n}}})}\\
&={\frac{1}{n}}{\color{black}{q^{(n)}}}{{{\color{black}{\ln}}}
{\frac{ {{\color{black}{u_{k}}}^{\alpha}}q^{(n)}{
{\color{black}{{\ddot{\rho}}_{n}}}}q^{(n)}}{ {q^{(n)}{{\bar{\rho}}_{\alpha}^{(n)}}q^{(n)}}}}}+{\frac{\alpha}{n}}{\color{black}{q^{(n)}}}{{\color{black}{\ln}}}
{\frac{1}{ A^{(n)}({\rho}^{\ast} || {\ddot{\rho}})}}\\
&={\frac{1}{n}}{\color{black}{q^{(n)}}}{{{\color{black}{\ln}}}
{\frac{q^{(n)}{{{\rho}^{\ast}}^{(n)}}q^{(n)}}{ {q^{(n)}{{\bar{\rho}}_{\alpha}^{(n)}}q^{(n)}}}}}+{\frac{\alpha}{n}}{\color{black}{q^{(n)}}}{{\color{black}{\ln}}}
{\frac{(\frac{ q^{(n)}
{\color{black}{{\ddot{\rho}}_{n}}}{q^{(n)}}}{q^{(n)}{{\rho}^{\ast}}^{(n)}q^{(n)}})^{1/{\alpha}}{\color{black}{u_{k}}}}{    
A^{(n)}({\rho}^{\ast} || {\ddot{\rho}})}}\\
&= {\frac{1}{n}}{\color{black}{q^{(n)}}}{{{\color{black}{\ln}}}
{\frac{q^{(n)}{{\rho}^{\ast}}^{(n)}q^{(n)}}{ q^{(n)}{\bar{\rho}}_{\alpha}^{(n)}q^{(n)}}}}+ {\frac{\alpha}{n}}{\color{black}{q^{(n)}}}{{\color{black}{\ln}}}
{\frac{(\frac{ q^{(n)}
{\color{black}{{\ddot{\rho}}_{n}}}{q^{(n)}}}{q^{(n)}{{\rho}^{\ast}}^{(n)}q^{(n)}})^{1-\lambda}{\color{black}{u_{k}}}}{    
A^{(n)}{({\rho}^{\ast} || {\ddot{\rho}})}}}\\
&\leq{\frac{1}{n}}{\color{black}{q^{(n)}}}{{{\color{black}{\ln}}}
{\frac{q^{(n)}{{\rho}^{\ast}}^{(n)}q^{(n)}}{ q^{(n)}{\bar{\rho}}_{\alpha}^{(n)}q^{(n)}}}}+ {\frac{\alpha}{n}}{\color{black}{q^{(n)}}}{{\color{black}{\ln}}}
{\sum_{{\color{black}{m\in{\mathbb{N}}}}}{\frac{(\frac{{q}^{(n)}
{\color{black}{{{\rho}}_{m}}}{{q}^{(n)}}}{{q}^{(n)}{{\rho}^{\ast}}^{(n)}{q}^{(n)}})^{1-\lambda}{\color{black}{u_{m}}}}{    
A^{(n)}{({\rho}^{\ast} || { {\color{black}{{\rho}}_{m}}  })}}}}\\
\text{Therefore,} &E_{{{\rho}^{\ast}}^{(n)}}[{\bar{d}}_{\lambda}{({{\rho}^{\ast}}^{(n)}\Vert
{\color{black}{{\ddot{\rho}}_{n}}})}]\leq(\frac{1}{n})S_{nat}({{\rho}^{\ast}}^{(n)} || {\bar{\rho}}_{\alpha}^{(n)})+ \\
&(\frac{\alpha}{n}) Tr({\color{black}{{{{\rho}^{\ast}}^{(n)}}}} {{\color{black}{\ln}}}
{\sum_{{\color{black}{m\in{\mathbb{N}}}}}{\frac{(\frac{\boldsymbol{q}^{(n)}
{\color{black}{{{\rho}}_{m}}}{\boldsymbol{q}^{(n)}}}{\boldsymbol{q}^{(n)}{{\rho}^{\ast}}^{(n)}\boldsymbol{q}^{(n)}})^{1-\lambda}{\color{black}{u_{m}}}}{    
A^{(n)}{({\rho}^{\ast} || {{\color{black}{\rho_m}}  })}}}}  )
\end{array}
\end{equation*}
{\color{black}{
where ${\boldsymbol{q}}^{(n)}$ is a random projection under the density matrix $\rho^{\ast}$ with values in $Q^{(n)}.$

Since $Q\in {{\pi}_{0}{(\mathbb H)}}$ and
$$Tr(q^{(n)}{\rho_{k}^{(n)}}q^{(n)}{{\color{black}{\ln}}}
{q^{(n)}{\rho_{l}^{(n)}}q^{(n)}})
= Tr(q^{(n)}{\rho_{k}^{(n)}}q^{(n)})Tr(q^{(n)}lnq^{(n)}{\rho}_{l}^{(n)}q^{(n)}),$$}}

By Jensen's inequality we have
\begin{equation*}
\begin{array}{rl}
&(\frac{\alpha}{n}) Tr({\color{black}{{{{\rho}^{\ast}}^{(n)}}}} {{\color{black}{\ln}}}
{\sum_{{\color{black}{m\in{\mathbb{N}}}}}{\frac{(\frac{ \boldsymbol{q}^{(n)}
{\color{black}{{{\rho}}_{m}}}{\boldsymbol{q}^{(n)}}}{\boldsymbol{q}^{(n)}{{\rho}^{\ast}}^{(n)}\boldsymbol{q}^{(n)}})^{1-\lambda}{\color{black}{u_{m}}}}{    
A^{(n)}{({\rho}^{\ast} || { {\color{black}{\rho_m}}  })}}}}  )\\
&\leq(\frac{\alpha}{n}) {{\color{black}{Tr\ln}}}
({\color{black}{{{{\rho}^{\ast}}^{(n)}}}}{\sum_{{\color{black}{m\in{\mathbb{N}}}}}{\frac{(\frac{\boldsymbol{q}^{(n)}
{\color{black}{{{\rho}}_{m}}}{\boldsymbol{q}^{(n)}}}{\boldsymbol{q}^{(n)}{{\rho}^{\ast}}^{(n)}\boldsymbol{q}^{(n)}})^{1-\lambda}{\color{black}{u_{m}}}}{    
A^{(n)}{({\rho}^{\ast} || { {\color{black}{\rho_m}}  })}}}}  )\\
&=(\frac{\alpha}{n}){{\color{black}{\ln}}}
{\sum_{{\color{black}{m\in{\mathbb{N}}}}}{[(\frac{{\color{black}{u_{m}}}}{{ A^{(n)}({\rho}^{\ast} || {\color{black}{\rho_m}})}})Tr(\sum_{q^{(n)}\in Q^{(n)}} ({{{q^{(n)}}{{\rho}^{\ast}}^{(n)}q^{(n)}}})^{\lambda}({ q^{(n)}{\color{black}{\rho_m}}^{(n)}{q^{(n)}}})^{1-\lambda})]}}\\
&=(\frac{\alpha}{n}){{\color{black}{\ln}}}
{\sum_{{\color{black}{m\in{\mathbb{N}}}}}[(\frac{{\color{black}{u_{m}}}}{{ A^{(n)}({\rho}^{\ast} || {\color{black}{\rho_m}})}})Tr({\otimes}^{n}({{\rho}^{\ast\lambda}}{\color{black}{\rho_m}}^{1-\lambda}))]}\\
&=(\frac{\alpha}{n}){{\color{black}{\ln}}}
{\sum_{{\color{black}{m\in{\mathbb{N}}}}}[(\frac{{\color{black}{u_{m}}}}{{ A^{(n)}({\rho}^{\ast} || {\color{black}{\rho_m}})}}){\Pi}^{n}Tr({{\rho}^{\ast\lambda}}{\color{black}{\rho_m}}^{1-\lambda})]}\\
&=(\frac{\alpha}{n}){{\color{black}{\ln}}}
{\sum_{{\color{black}{m\in{\mathbb{N}}}}}[(\frac{{\color{black}{u_{m}}}}{{ A^{(n)}({\rho}^{\ast} || {\color{black}{\rho_m}})}}) {{ A^{(n)}({\rho}^{\ast} || {\color{black}{\rho_m}})}}]}\\
&=(\frac{\alpha}{n}){{\color{black}{\ln}}}
{\sum_{{\color{black}{m\in{\mathbb{N}}}}}{{\color{black}{u_{m}}}}}.
\end{array}
\end{equation*}

But $ \sum_{{\color{black}{m\in{\mathbb N}}}}{ }{\color{black}{u_{m}}}{=1}.$ Therefore,
$E_{{\rho}^{\ast}}[{\bar{d}}_{\lambda}{({\rho}^{(n)}\Vert \ddot{\rho}_{n}^{(n)})}]\leq(\frac{1}{n})S_{nat}({{\rho}^{\ast}}^{(n)} || {\bar{\rho}}_{\alpha}^{(n)}).$

\begin{corollary}
From the above theorem, Definition 9 and the relation between Renyi divergences and Hellinger distance explained above we have:
\begin{enumerate}
\item $lim_{n\rightarrow {\infty}}{E_{\rho^{\ast(n)}_{Q}}{({{He}^{2}{(\rho^{\ast(n)}_{Q}\Vert \ddot{\rho}_{ nQ}^{(n)})}})}}=0.$
\item Let $\alpha>1$ and $0<\lambda={1-1/{\alpha}}$. Then,
$$lim_{n\rightarrow {\infty}}{ E_{\rho^{\ast(n)}_{Q}}{({\bar{d}}_{\lambda}{(\rho^{\ast(n)}_{Q}\Vert \ddot{\rho}_{nQ}^{(n)})})}}=0.$$
\end{enumerate}

\end{corollary}

\section{Applications}
As we described before, estimation and prediction are the most important purposes of quantum statistical inference and particularly this paper. In order to show the advantages of our method, in this section we explain the usage of this method by two examples. The first example that we choose is selecting a density matrix among three ones which are originally considered in [5]. For multiple ions quantum tomography, two famous traditional methods, the Akaike information criterion (AIC) and the Bayesian information criterion (BIC) are used for estimation. For more information please see [5].

In this case, the quantum model consists of three one-ion states of different degrees of purity: a pure state, one with eigenvalues (0.95, 0.05), and the other with eigenvalues (0.72, 0.28). For each state, they simulated data sets with varying numbers of repetitions n = 10, 50, 100, 250, 500. Table 1, shows the number of times (out of 1000 samples) that BIC and AIC chose correctly, [5].
\begin{table}
\centering
  \begin{threeparttable}
    \caption{AIC and BIC Model Selection}
\begin{tabular}{cc|c|c|c|c|l|}
\cline{3-7}
& & \multicolumn{5}{ c| }{Measurement Repetition} \\ \cline{3-7}
& & 10 & 50 & 100 & 250 & 500\\ \cline{1-7}
\multicolumn{1}{ |c  }{\multirow{2}{*}{State 1} } &
\multicolumn{1}{ |c| }{BIC} & 987 & 990 & 994 & 992 & 996    \\ \cline{2-7}
\multicolumn{1}{ |c  }{}                        &
\multicolumn{1}{ |c| }{AIC} & 945 & 944 & 919 & 927 & 930    \\ \cline{1-7}
\multicolumn{1}{ |c  }{\multirow{2}{*}{State 2} } &
\multicolumn{1}{ |c| }{BIC} & 25 & 83 & 183 & 394 & 706 \\ \cline{2-7}
\multicolumn{1}{ |c  }{}                        &
\multicolumn{1}{ |c| }{AIC} & 77 & 312 & 502 & 802 & 942 \\ \cline{1-7}
\multicolumn{1}{ |c  }{\multirow{2}{*}{State 3} } &
\multicolumn{1}{ |c| }{BIC} & 384 & 973 & 998 & 997 & 988 \\ \cline{2-7}
\multicolumn{1}{ |c  }{}                        &
\multicolumn{1}{ |c| }{AIC} & 594 & 992 & 998 & 997 & 998 \\ \cline{1-7}
\end{tabular}
    \begin{tablenotes}
      \small
      \item  Performance of BIC and AIC model selection for 3 states: pure (state 1), almost pure (state 2), and mixed (state 3). This table is based on the results in [5].
    \end{tablenotes}
  \end{threeparttable}
\end{table}

Now we choose among these states with the quantum version of classical two-part code estimation, semi-density matrices.

Let $$\rho= \begin{bmatrix}
a & b \\
b & 1-a \\
\end{bmatrix},Q=\{\vert 0\rangle\langle 0\vert,\vert 1\rangle\langle 1\vert\}.$$

Let $q_{i_j}=\vert 0\rangle\langle 0\vert$ then $q_{i_j}\rho q_{i_j}=a q_{i_j}$ and if $q_{i_j}=\vert 1\rangle\langle 1\vert$ then $q_{i_j}\rho q_{i_j}=(1-a) q_{i_j}$, and

$$q^{(n)}_{I}(2^{-L(\rho)} \rho^{(n)})q^{(n)}_{I}=2^{-L(\rho)}(q_{i_1}\rho q_{i_1})\otimes (q_{i_2}\rho q_{i_2})\otimes\cdots\otimes (q_{i_n}\rho q_{i_n}).$$

Assume that $q^{(n)}_I\in Q^{(n)}$ consists of $k-$times $\vert 0\rangle\langle 0\vert$ and $(n-k)-$times $\vert 1\rangle\langle 1\vert.$ Then:

$$q^{(n)}_{I}(2^{-L(\rho)} \rho^{(n)})q^{(n)}_{I}=2^{-L(\rho)} a^{k}{(1-a)}^{(n-k)} q^{(n)}_{I}$$

Now let us calculate this for the states considered in [5].

\begin{example}
\begin{enumerate}
\item For the states in [5], we define the following quantum generalized model
$$\mathcal{M}=\bigg\{\rho_1=
\dfrac{1}{3}\begin{bmatrix}
1 & 0 \\
0 & 0 \\
\end{bmatrix},\rho_2=
\dfrac{4}{9}\begin{bmatrix}
0.95 & 0 \\
0 & 0.05 \\
\end{bmatrix},\rho_3=
\dfrac{2}{9}\begin{bmatrix}
0.72 & 0 \\
0 & 0.28\\
\end{bmatrix}\bigg\}$$

\[q^{(n)}_{I}(2^{-L(\rho_1)} \rho^{(n)}_1)q^{(n)}_{I} = \left\{
  \begin{array}{lr}
    \dfrac{1}{3}q^{(n)}_{I} &  k=n\\
    0 &  \text{Otherwise}
  \end{array}
\right.
\]

$$q^{(n)}_{I}(2^{-L(\rho_2)} \rho^{(n)}_2)q^{(n)}_{I}=\dfrac{4}{9}\big(0.95\big)^k\big(0.05\big)^{(n-k)}q^{(n)}_{I}$$

$$q^{(n)}_{I}(2^{-L(\rho_3)} \rho^{(n)}_3)q^{(n)}_{I}=\dfrac{2}{9}\big(0.72\big)^k\big(0.28\big)^{(n-k)}q^{(n)}_{I}$$

For each state, we simulated datasets with varying numbers of repetitions n = 10, 50, 100, 250, 500. Table 2, shows the number of times (out of 1000 samples) that the quantum version of classical two-part code estimation chose correctly.
\begin{table}
\centering
  \begin{threeparttable}
    \caption{The quantum version of classical two-part code estimation}
\begin{tabular}{cc|c|c|c|c|l|}
\cline{3-7}
& & \multicolumn{5}{ c| }{Measurement Repetition} \\ \cline{3-7}
& & 10 & 50 & 100 & 250 & 500\\ \cline{2-7}
\multicolumn{1}{ c  }{} &
\multicolumn{1}{ |c| }{State 1} & 1000 & 1000 & 1000 & 1000 &   1000  \\ \cline{2-7}
\multicolumn{1}{ c  }{}                        &
\multicolumn{1}{ |c| }{State 2} & 336 & 926 & 995 & 1000 &   1000  \\ \cline{2-7}
\multicolumn{1}{ c  }{} &
\multicolumn{1}{ |c| }{State 3} & 747 & 980 & 998 & 1000 & 1000 \\ \cline{2-7}
\end{tabular}
    \begin{tablenotes}
      \small
      \item  Performance of the quantum version of classical two-part code estimation for 3 states: $\rho_1$ (state 1), $\rho_2$ (state 2), and $\rho_3$ (state 3).
    \end{tablenotes}
  \end{threeparttable}
\end{table}

As expected, for small sample sizes, n, the quantum version of classical two-part code estimation may select the wrong model because it has a built-in preference for “simple” models. But for all large n, it will select the correct model. Yet for the small $n$, it is far better than classical methods, like AIC and BIC. In the case of the pure state because of the appropriate choice of weight, it never missed and always chose correctly. On the other hand, it avoids overfitting and it did well for the mixed states too. AIC and BIC have mistakes even for the large number of $n$. The comparison between Table 1 and Table 2 will show the difference between using semi-density matrices and common traditional models.

\item If we use the following quantum generalized model,

$$\mathcal{M}=\bigg\{\rho_1=
\dfrac{4}{9}\begin{bmatrix}
1 & 0 \\
0 & 0 \\
\end{bmatrix},\rho_2=
\dfrac{1}{3}\begin{bmatrix}
0.95 & 0 \\
0 & 0.05 \\
\end{bmatrix},\rho_3=
\dfrac{2}{9}\begin{bmatrix}
0.72 & 0 \\
0 & 0.28\\
\end{bmatrix}\bigg\}$$

that the weights have an inverse relationship with the Shannon entropy of each state, the result will be as follows,

\[q^{(n)}_{I}(2^{-L(\rho_1)} \rho^{(n)}_1)q^{(n)}_{I} = \left\{
  \begin{array}{lr}
    \dfrac{4}{9}q^{(n)}_{I} &  k=n\\
    0 &  \text{Otherwise}
  \end{array}
\right.
\]

$$q^{(n)}_{I}(2^{-L(\rho_2)} \rho^{(n)}_2)q^{(n)}_{I}=\dfrac{1}{3}\big(0.95\big)^k\big(0.05\big)^{(n-k)}q^{(n)}_{I}$$

$$q^{(n)}_{I}(2^{-L(\rho_3)} \rho^{(n)}_3)q^{(n)}_{I}=\dfrac{2}{9}\big(0.72\big)^k\big(0.28\big)^{(n-k)}q^{(n)}_{I}$$

For each state, we simulated datasets with varying numbers of repetitions n = 10, 50, 100, 250, 500. Table 2, shows the number of times (out of 1000 samples) that the quantum version of classical two-part code estimation chose correctly.
\begin{table}
\centering
  \begin{threeparttable}
    \caption{The quantum version of classical two-part code estimation}
\begin{tabular}{cc|c|c|c|c|l|}
\cline{3-7}
& & \multicolumn{5}{ c| }{Measurement Repetition} \\ \cline{3-7}
& & 10 & 50 & 100 & 250 & 500\\ \cline{2-7}
\multicolumn{1}{ c  }{} &
\multicolumn{1}{ |c| }{State 1} & 1000 & 1000 & 1000 & 1000 &   1000  \\ \cline{2-7}
\multicolumn{1}{ c  }{}                        &
\multicolumn{1}{ |c| }{State 2} & 345 & 904 & 995 & 1000 &   1000  \\ \cline{2-7}
\multicolumn{1}{ c  }{} &
\multicolumn{1}{ |c| }{State 3} & 732 & 981 & 995 & 1000 & 1000 \\ \cline{2-7}
\end{tabular}
    \begin{tablenotes}
      \small
      \item  Performance of the quantum version of classical two-part code estimation for 3 states: $\rho_1$ (state 1), $\rho_2$ (state 2), and $\rho_3$ (state 3).
    \end{tablenotes}
  \end{threeparttable}
\end{table}
\item In the last part of this example, let us calculate the quantum version of classical two-part code for sequences of length n=10 and n=50 generated by an unknown model. We calculate $q^{(n)}_{I}(2^{-L(\rho_i)} \rho^{(n)}_i)q^{(n)}_{I}$ to observe which of the above model is the best fit for generating this sequence based on the quantum version of classical two-part code.

$n=10,$
$D=\vert 0010100110\rangle$

\[q^{(n)}_{I}(2^{-L(\rho_1)} \rho^{(n)}_1)q^{(n)}_{I} =  0
\]

$q^{(n)}_{I}(2^{-L(\rho_2)} \rho^{(n)}_2)q^{(n)}_{I}=\dfrac{1}{3}\big(0.95\big)^6\big(0.05\big)^{4}q^{(n)}_{I}=0.000001531441 q^{(n)}_{I}$

$q^{(n)}_{I}(2^{-L(\rho_3)} \rho^{(n)}_3)q^{(n)}_{I}=\dfrac{2}{9}\big(0.72\big)^6\big(0.28\big)^{4}q^{(n)}_{I}=0.0001544952 q^{(n)}_{I}$

now we continue generating and make a sequence with length 50 to see how the calculation differs.

$n=50$
$D=\vert 00101001100110101101101001010101010100101010100010\rangle$

\[q^{(n)}_{I}(2^{-L(\rho_1)} \rho^{(n)}_1)q^{(n)}_{I} =  0
\]

$q^{(n)}_{I}(2^{-L(\rho_2)} \rho^{(n)}_2)q^{(n)}_{I}=\dfrac{1}{3}\big(0.95\big)^27\big(0.05\big)^{23}q^{(n)}_{I}=9.94778\times 10^{-32} q^{(n)}_{I}$

$q^{(n)}_{I}(2^{-L(\rho_3)} \rho^{(n)}_3)q^{(n)}_{I}=\dfrac{2}{9}\big(0.72\big)^27\big(0.28\big)^{23} q^{(n)}_{I}=1.336786\times 10^{-18} q^{(n)}_{I}$

Although the sequence is not necessarily generated by one of the above models, the closest model for generating sequence $D$ is the third semi-density matrix.

\end{enumerate}

\end{example}

In the next example, we will show a concrete example of calculating a universal quantum source and predicting the $n+1$-th outcome by a quantum strategy.

\begin{example}
Let $\mathcal{M}$ be the following quantum generalized model
$$\mathcal{M}=\bigg\{\rho_1=
\dfrac{1}{12}\begin{bmatrix}
\dfrac{1}{2} & \dfrac{1}{2} \\
\dfrac{1}{2} & \dfrac{1}{2} \\
\end{bmatrix},\rho_2=
\dfrac{4}{12}\begin{bmatrix}
\dfrac{2}{3} & \dfrac{2}{9} \\
\dfrac{2}{9} & \dfrac{1}{3} \\
\end{bmatrix},\rho_3=
\dfrac{7}{12}\begin{bmatrix}
\dfrac{1}{4} & \dfrac{1}{8} \\
\dfrac{1}{8} & \dfrac{3}{4}\\
\end{bmatrix}\bigg\},$$
and we want to predict the $n+1$-th outcome, after observing $n$ measurements. Based on what we said in the previous sections the $Q-$universal quantum source is as follows:

$q^{(n)}_{I}\bar{\rho}^{(n)}q^{(n)}_{I}=\bigg(\dfrac{1}{12}\bigg(\dfrac{1}{2}\bigg)^k\bigg(\dfrac{1}{2}\bigg)^{(n-k)}+\dfrac{4}{12}\bigg(\dfrac{2}{3}\bigg)^k\bigg(\dfrac{1}{3}\bigg)^{(n-k)}+\dfrac{7}{12}\bigg(\dfrac{1}{4}\bigg)^k\bigg(\dfrac{3}{4}\bigg)^{(n-k)}\bigg)q^{(n)}_{I}.$
\\

The quantum strategy associated with the above universal model is

\begin{equation*}
\begin{array}{rl}
&\hat{\rho}^{n+1}(\vert 0\rangle\langle 0\vert \big |q^{(n)}_{I})=\\
&\\
&\dfrac{\dfrac{1}{12}\bigg(\dfrac{1}{2}\bigg)^{k+1}\bigg(\dfrac{1}{2}\bigg)^{(n-k)}+\dfrac{4}{12}\bigg(\dfrac{2}{3}\bigg)^{k+1}\bigg(\dfrac{1}{3}\bigg)^{(n-k)}+\dfrac{7}{12}\bigg(\dfrac{1}{4}\bigg)^{k+1}\bigg(\dfrac{3}{4}\bigg)^{(n-k)}}{\dfrac{1}{12}\bigg(\dfrac{1}{2}\bigg)^k\bigg(\dfrac{1}{2}\bigg)^{(n-k)}+\dfrac{4}{12}\bigg(\dfrac{2}{3}\bigg)^k\bigg(\dfrac{1}{3}\bigg)^{(n-k)}+\dfrac{7}{12}\bigg(\dfrac{1}{4}\bigg)^k\bigg(\dfrac{3}{4}\bigg)^{(n-k)}}\vert 0\rangle\langle 0\vert=
\end{array}
\end{equation*}
$$\dfrac{1}{12}\times \dfrac{6^{n+1}+2^{2n+k+5}+7\times 3^{2n-k-1}}{6^{n}+2^{2n+k+2}+7\times 3^{2n-k}}\vert 0\rangle\langle 0\vert.$$

Therefore the probability of $\vert 0\rangle\langle 0\vert$ given $q_I^{(n)}$ is

$$\dfrac{1}{12}\times \dfrac{6^{n+1}+2^{2n+k+5}+7\times 3^{2n-k-1}}{6^{n}+2^{2n+k+2}+7\times 3^{2n-k}}$$

\end{example}

\section*{Acknowledgment}
The authors would like to express their very great appreciation to Prof. Fabio Benatti for his valuable and constructive suggestions. His willingness to give his time so generously has been very much appreciated. We would also like to thank Prof. Peter D. Grunewald for his very interesting book, ``the MDL principle'', from which we learned the classical MDL principle.

\end{document}